\documentclass[%
 aip,
 amsmath,amssymb,
 reprint,%
]{revtex4-1}

\usepackage{graphicx}
\usepackage{dcolumn}
\usepackage{bm}

\usepackage[utf8]{inputenc}
\usepackage[T1]{fontenc}
\usepackage{mathptmx}

\usepackage{vale}
\usepackage{mathtools}
\usepackage[varg]{txfonts}   
\usepackage{color}
\usepackage{colortbl}
\usepackage{array}
\usepackage{amssymb}
\usepackage{amsmath}
\usepackage{mathtools}
\usepackage{tabularx}
\usepackage{multirow}
\usepackage{subfigure}
\usepackage{psfrag}
\usepackage{paralist}
\usepackage{setspace}
\usepackage{comment}
\usepackage{morefloats}
\usepackage{vale}
\usepackage{natbib}
\usepackage{bm}

\begin{document}

\preprint{AIP/123-QED}

\title{Eulerian--Lagrangian modelling of bio--aerosols irradiated by UV--C light in relation to SARS--CoV--2 transmission}

\author{V. D'Alessandro}
 \email{v.dalessandro@univpm.it.}
\affiliation{Dipartimento di Ingegneria Industriale e Scienze Matematiche, Università Politecnica delle Marche, Via Brecce Bianche 12, 60131 Ancona (AN), Italy}

\author{M. Falone}%
\affiliation{Dipartimento di Ingegneria Industriale e Scienze Matematiche, Università Politecnica delle Marche, Via Brecce Bianche 12, 60131 Ancona (AN), Italy}

\author{L. Giammichele}%
\affiliation{Dipartimento di Ingegneria Industriale e Scienze Matematiche, Università Politecnica delle Marche, Via Brecce Bianche 12, 60131 Ancona (AN), Italy}

\author{R. Ricci}%
\affiliation{Dipartimento di Ingegneria Industriale e Scienze Matematiche, Università Politecnica delle Marche, Via Brecce Bianche 12, 60131 Ancona (AN), Italy}

\date{\today}

\begin{abstract}
It is well known that several viruses, as well as SARS--CoV--2, can be transmitted through airborne
diffusion of saliva micro-droplets. For this reason many reserach groups have been
devoted their efforts in order to gain new insight into the transport of fluids and particles
originted from human respiratory tracts.\\
This paper aims to provide a contribution to the numerical modelling of bio--aerosols. 
In particular, the well--known problem around the safety distance to be held for avoiding virus transmission 
in the absence of external wind is further investigated. Thus, new indexes capable of evaluating the 
contamination risk are introduced and the possibility to inactivate virus particles by means of an external 
UV--C radiation source is studied. For this purpose, a new model
which takes into account biological inactivation deriving from UV--C exposure in a Eulerian--Lagrangian 
framework is presented.
\end{abstract}

\maketitle

%

\section{Introduction}\label{sec:intro}
As largely reported in the open literature, several viruses, 
as well as SARS--CoV--2, can be transmitted through airborne 
diffusion of saliva micro-droplets\cite{Siegel:2007} too small to 
be seen with the naked eye. Typical infection mechanisms are the following:
\begin{inparaenum}[(i)]
\item direct transfer of large droplets expelled at high momentum to the receiver's conjunctiva, mouth, or nose;
\item physical contact with droplets deposited on the surface and subsequent absorption to the nasal
mucosa of the receiver;
\item inhalation by the recipient of respiratory ejected aerosolised droplet nuclei
\end{inparaenum}\cite{Mittal2020}.
%
For this reason, many countries in the world have imposed variable social distances to be kept between persons. 
This restriction has been adopted since the safety distance must be guaranteed in order to allow the most elevate 
number of the droplets to fall down and reach the floor, or even evaporate, after their emission from a mouth or nose. 
Hence, it is straightforward to understand that the correct and rigorous study of saliva droplets dynamics, involving 
all the relevant biological and physical phenomena, is the key ingredient to determine the guidelines on social distancing, 
face masks wearing as well as the implementation of new practices in the daily social life. 
It is also worth noting that the physical phenomena involved in the droplets transmission process are very complex. 
Indeed, after their emission, micro--droplets travel as results of their inertia and their aerodynamic interaction with air. 
Moreover, the mass of the droplet can vary due to evaporation which is strictly connected to air temperature and relative humidity.\\
After SARS epidemic, that started at end of 2002, several studies about airborne droplets transmission have been
published in medical and non--medical journals.
As already introduced, both computational and experimental models have been employed by investigators with particular emphasis on indoor 
conditions\cite{Melikov:2018}.
Some research groups have carried on chamber experiments. However, an essential disadvantage of this kind of approaches is that traditional 
measurements are too discrete,~\emph{i.e.} only a few points can be investigated at the same time. Although laser--based techniques, 
such as Particle Image Velocimetry (PIV), allow measuring 2D or even 3D velocity fields, they cannot provide a 
quantitative evaluation of the cross--infection risks\cite{LICINA}.
Besides, flow and concentration sensors can produce significant disturbances for droplet's transport inside the mouth/nose zone. 
Lastly, it is worth noting that similar apparatuses are expensive. Numerical modelling can be considered a valid alternative to overcome
these limitations\cite{NIELSEN}.
Flow fields and droplet dynamics can be computed with a very high temporal resolution, far less
than the scale of the human breathing activities. Moreover, computer simulations have more considerable flexibility than experimental 
investigations. However, an important issue for CFD simulations is the obtained accuracy, which is influenced by geometrical 
simplifications as well as the failure of adopted models\cite{NIELSEN}.\\
After unprecedented COVID--19 pandemic several research groups have been 
devoted their efforts in order to gain new insight into the transport of fluids and particles 
emanating from human respiratory tracts. With this work we want to contribute to this emerging and 
important research field in the numerical modelling context.
In this area we can find the Vuorinen et al.~\cite{VUORINEN2020104866} paper
which discuss the physical processes related to the aerosolisation of the exhaled droplets
by means of Large--Eddy--Simulation (LES). A similar approach was used by Pendar and Pascoa~\cite{Pendar}
which focused their effort on the development of reliable model for the emission of saliva droplets
during coughing and sneezing.
Differently, Dbouk and Drikakis~\cite{Dbouk, Dbouk1, Dbouk2} developed a Eulerian--Lagrangian
model based on Reynolds--Averaged Navier--Stokes (RANS) equations for the simulation of human cough;
the impact of face masks and  weather conditions on the droplet evaporation phenomenon were studied.
Also Busco et al.~\cite{Busco} adopted RANS equations to model the carrier fluid in sneezing and 
asymptomatic conditions. In Abuhegazy et al.~\cite{Abu_pof} a RANS based Eulerian--Lagrangian model
of bio--aerosol transport in a classroom, with relevance to COVID--19, is presented. 
Lastly, Li et al.~\cite{Li_pof} presented a CFD model for droplets evaporation
and transport in tropical outdoor environment.\\
In this paper, a new computational model, relying on the well established OpenFOAM library~\cite{OF_paper}, for the evaluation
of saliva droplets' dynamics during coughing is presented. Starting from the work published in the more relevant 
papers (above briefly discussed), the authors intend to provide CFD practitioners with several crucial data about 
case settings. Also, two new indexes are introduced in order to evaluate contamination risk. Lastly, a focus on 
the possibility to reduce SARS--CoV--2 transmission potential by means of UV--C radiation is shown. 
This topic was already investigated by Buchan et al.\cite{Buchan2020}. In the cited paper, the UV--C effect 
was assessed considering saliva droplets in a dilute solution with air. On the contrary, the present work aims 
to introduce an approach capable of including the biological inactivation related to UV--C field in an 
Eulerian--Lagrangian framework.\\
This paper is organised as follows: the governing equations are presented in Section~\ref{sec:goveq}, while
the numerical discretisation techniques are discussed in Section~\ref{sec:num}.
Numerical results are shown in Section~\ref{sec:results}. Lastly, Section~\ref{sec:concl} contains the conclusions.

\section{Governing equations}
\label{sec:goveq}
Numerical simulations are developed using an Eulerian--Lagrangian
framework in which Eulerian approach is applied to the atmospheric 
air. The Lagrangian reference frame is adopted for dispersed droplets generated
by breathing.

\subsection{Eulerian phase}
For Eulerian phase compressible RANS equations are used:
\begin{equation} \label{eq:ns_v_cons}
  \begin{aligned}
    & \frac{\partial \rhobar}{\partial t} + 
      \frac{\partial  }{\partial x_j} \left( \rhobar \utilde_j \right)   = s_m, \\ 
    & \frac{\partial} {\partial t} \left( \rhobar \utilde_i \right)   + 
      \frac{\partial}{\partial x_j}  \left(  \rhobar \utilde_i \utilde_j   \right)   = -\frac{\partial \overline{p}}{\partial x_i} + \frac{\partial \hat{\tau}_{ij}}{\partial x_j} +  \rhobar g \delta_{i3}  +  s_{m,i}, \\ 
    & \frac{\partial}{\partial t} \left(  \rhobar \tilde{E}   \right)  + 
      \frac{\partial}{\partial x_j} \left( \rhobar \utilde_j \tilde{H}   \right)   =  -\frac{\partial q_j}{\partial x_j}   +\frac{\partial }{\partial x_j} \left( \utilde_i \hat{\tau}_{ij}   \right) 
      + s_{e},\\ 
    & \frac{\partial }{\partial t} \left( \rhobar \tilde{Y}_k   \right)  + 
      \frac{\partial }{\partial x_j} \left( \rhobar \utilde_j \tilde{Y}_k  \right)   = -\frac{\partial m_{k,j}}{\partial x_j}    + s_{Y_k},\\ 
  \end{aligned}
\end{equation}
where $\rhobar$, $\utilde_i$, $\overline{p}$, $\tilde{T}$ and ${\tilde{Y_k}}$
denote density, velocity component in $x_i$ direction, pressure, temperature and chemical specie
$k$ mass fraction. $\tilde{E}$ and $\tilde{H}$ are, respectively, the total internal energy and entalpy.
Note that the overbar and the tilde are filtering operators which are introduced
for unweighted and density--weighted averages.\\
%
The unclosed terms reported in eq.~\ref{eq:ns_v_cons} are handled 
as follows:
\begin{equation} \label{eq:uncl}
  \begin{aligned}
     &  q_j = -c_p \left( \frac{\mu}{\mathrm{Pr}} + \frac{\mu_t}{\mathrm{Pr_t}} \right)   \frac{ \partial \tilde{T}}{\partial x_j} , \\ 
     &  m_{k,j} = -\rhobar  \left( \frac{\mu}{\mathrm{Sc_k}} + \frac{\mu_t}{\mathrm{Sc_{k,t}}} \right)\frac{ \partial \tilde{Y}_k}{\partial x_j} . 
  \end{aligned}
\end{equation}
in eq.~\ref{eq:uncl} the symbol $c_p$ represents the specific heat at constant pressure and $\mu$ is the viscosity. 
$\mathrm{Pr}$ and $\mathrm{Sc_k}$ are molecular Prandtl and 
Schmidt numbers, the lower-script $t$ indicates the turbulent version of the previous dimensionless groups. 
The stress tensor $\hat{\tau}_{ij}$ is evaluated as follows:
\begin{equation}\label{eq:stress}
\hat{\tau}_{ij} = 2 \mu \left( \tilde{S}_{ij} - \frac{1}{3} \frac{\partial \utilde_k}{\partial x_k} \delta_{ij} \right) + \tau_{ij} 
\end{equation}
where
\begin{equation}\label{eq:stress2}
{\tau}_{ij} = 2 \mu_t \left( \tilde{S}_{ij} - \frac{1}{3}\frac{\partial \utilde_k}{\partial x_k} \delta_{ij} \right) - \frac{2}{3} \rhobar \overline{k} \delta_{ij}, 
\end{equation}

\noindent $\overline{k}$ is the average turbulent kinetic energy and  $\tilde{S}_{ij}$ the strain--rate tensor.
Turbulence modelling is performed using standard SST $k$--$\omega$, developed by Menter\cite{Menter.1994}, not described here for  
compactness. Polynomial equation of state was adopted and polynomial correlations were used for thermophysical properties.\\
The source terms $s_m$, $s_{m,i}$,  $s_{e}$ and $s_{Y_k}$  correspond to coupling 
between Lagrangian and Eulerian phases with respect to mass, momentum, energy and species, respectively.
The particle--source--in--cell (PSI--Cell) method~\cite{Crowe:1977} for source terms 
manipulation is adopted.
\subsection{Lagrangian phase}
Saliva droplets are tracked using a Lagrangian frame throughout the computational domain.
It is crucial to put in evidence that, within OpenFOAM Lagrangian libraries,
for efficiency reasons, the concept of computational parcel is adopted. The droplets
are organised in groups and each parcel represents 
the centre of mass of a small cloud of droplets having the same properties.  
Assuming non--collisional spherical parcels, position and velocity are the results 
of the trajectory and momentum equations:
\begin{equation} \label{eq:lpt1}
  \begin{aligned}
& \frac{d {\bf x}_{P,i}}{d t} = {\bf u}_{P,i} , \\
& m_{P,i}\frac{d {\bf u}_{P,i}}{d t} = {\bf F}^G_{P,i}  + {\bf F}^D_{P,i}  , \\
  \end{aligned}
\end{equation}
with parcel velocity ${\bf u}_{P,i}$, mass $m_{P,i}$ and position ${{\bf x}_{P,i}}$.
The forces acting on the generic $i$--th parcel in eq.~\ref{eq:lpt1}, are identifiable with two contributions:
gravity force ( ${\bf F}^G_{P,i}$  ) and aerodynamic drag force  (${\bf F}^D_{P,i}$).
Gravity force takes also into account buoyancy in the following way:
\begin{equation} \label{eq:lpt_g}
{\bf F}^G_{P,i} = m_{P,i} {\bf g} \left( 1- \frac{\overline{\rho}}{\rho_P}\right),
\end{equation}
where ${\rho_P}$ is the density of the generic element of the discrete phase. The aerodynamic drag force, ${\bf F}^D_{P,i}$, is :
\begin{equation} \label{eq:lpt_d}
{\bf F}^D_{P,i} = \rhobar C_D \frac{\pi D_P^2}{8} \left( \tilde{{\bf u}}  - {\bf u}_{P,i} \right)  \left| \tilde{{\bf u}}  - {\bf u}_{P,i} \right| ,
\end{equation}
the drag coefficient, $C_D$, is evaluated from a correlation based on Putnam~\cite{Putnam:1961} paper being then $D_P$ the particles' diameter.
Additional forces including the following components: pressure, virtual mass, Basset and Brownian, 
are not included as done by other authors~\cite{Busco, Abu_pof, Li_pof}.
As a matter of the fact, the particles considered in the present work are sufficiently small to neglect pressure 
and virtual mass forces and sufficiently large to neglect Brownian force~\cite{Abu_pof, aerosol:2007, ZHAO20041}.
This evidence held true also in the preliminary computations carried out in this research.\\
%
%
%
The mass conservation equation reads:
\begin{equation} \label{eq:mass}
 \frac{d m_{P,i} }{d t} = -\dot{m}^{ev}_{P,i} , \\
\end{equation}
where the evaporation term, $\dot{m}^{ev}_{P,i}$, is governed by the diffusive flux of vapor and
the mass transfer coefficient is obtained from the well established Ranz--Marshall correlation~\cite{Ranz:1952}.\\
Lastly, the parcel temperature, $T_{P,i}$, is obtained through the analytic solution
of the energy equation:
\begin{equation} \label{eq:lpt_egy}
m_{P,i} c_{p,i} \frac{d T_{P,i}}{d t} = h A_{P,i} \left( T_{P,i} - \tilde{T}  \right) + \dot{Q}_{ev}.
\end{equation}
In eq.~\ref{eq:lpt_egy} the convective heat transfer coefficient, $h$, is obtained from
the Ranz--Marshall correlation~\cite{Ranz:1952} for $\mathrm{Nu}$ number; $\dot{Q}_{ev}$ 
is the term including the heat transfer between continuous and discrete phase due to droplets
evaporation.\\
The Rosin--Rammler distribution\cite{Mugele1951} is used
for representing initial parcels' diameter:
\begin{equation} \label{eq:rosin}
f = \frac{n}{ \overline{D}_P} \left(  \frac{ D_P }{ \overline{D}_P}  \right)^{n-1}   \exp \left[  -  \left(  \frac{ D_P }{ \overline{D}_P}  \right)^n   \right]    
\end{equation}
in eq.~\ref{eq:rosin} we fix $n=8$ and $\overline{D}_p = 80$~$\mu m$ as in 
Dbouk and Drikakis~\cite{Dbouk1} who performed a fit of Xie et al.\cite{Xie:2009} experimental data regarding
human cough.  
The minimum diameter of the injected parcels is $10$~$\mu m$, while the maximum one is  $280$~$\mu m$. 
\subsection{UV--C inactivation modelling}
The present work shows a new model able to take into account
the presence of virus/bacterial particles in a bio--aerosol and
evaluate their biological inactivation produced by an external
UV--C field.\\
This is a complex multi-physics problem, and it was addressed in the available literature
by solving a transport equation for virus concentration~\cite{Beggs2000141,NOAKES2004489, Buchan2020}. 
Similar approaches can be considered appropriate for handling dilute solutions, but they are not suitable 
to investigate the interaction of UV--C light with a cloud of saliva droplets produced during cough or sneeze.\\
In the here described approach the number of active particles in each parcel, $N_{a,i}$, 
is estimated starting from the number of particles grouped inside the parcel
itself, $N_{p,i}$, as follows:
%
\begin{equation} \label{eq:chick_wat}
N_{a,i} = N_{p,i} - I_{a,i} 
\end{equation}
where
\begin{equation} \label{eq:chick_inact}
  \begin{aligned}
   I_{a,i} \left( t^{\left( k  \right)}  \right) &= \sum_{k=1}^{N_{ts}} N_{a,i} \left( t^{\left( k-1  \right)}  \right) F_{a,i}     \\
   F_{a,i} &=  1 - e^{-Z E_p \left(  {\bf x}_{P,i} ,  t^{\left( k  \right)}    \right) \Delta t} . 
  \end{aligned}
\end{equation}
In eq.~\ref{eq:chick_inact} the term $I_{a,i}$ is the number of particles inactivated by UV--C radiation in the parcel focusing on the point ${\bf x}_{P,i}$;
the integer parameter, $N_{ts}$, represents the current time--step index.\\
It is important to remark that the inactivation coefficient, $F_{a,i}$, is derived from
first-order Chick--Watson kinetics:
\begin{equation} \label{eq:chick_wat1}
\frac{N\left(t \right)}{N_0} = e^{-Z E_p t}. 
\end{equation}
In eq.~\ref{eq:chick_wat1} $N\left( t \right)$ and $N_0$ represent the number of active particles at the generic
time instant $t$ and $t=0$, respectively.
Differently, $Z$ is a susceptibility constant for the microorganism and $E_p$ is the mean irradiance of the UV--C field.
In this research work, we fix $Z=8.5281 \cdot 10^{-2}$~$m^{2}/J$
which is the average experimental value\cite{KowalskiUVC} obtained for a UV--C light ($\lambda = 254$~$nm$)
irradiating SARS--CoV--2.\\ 
As regards $E_p$, its estimation is achieved by means of the  
thermal radiation view factors method\cite{Modest}.
This technique was chosen for its capability
in well describing the intensity field due to cylindrical UV--lamps\cite{Kowalski2000}.
The fraction of the total radiation intensity emitted, that is collected by a parcel perpendicular to the lamp axis
and located in correspondence of its edge, is given by:
\begin{equation}\label{eq:viewF}
\begin{aligned}
F = \frac{L_l}{\pi H_l} \left[ 
\frac{1}{L_l}\arctan{\left(\frac{L_l}{\sqrt{H_l^2-1}}\right)} 
- \arctan {(M)} \right.\\
\left.+\frac{X-2H_l}{\sqrt{XY}} \arctan{\left(M \sqrt{\frac{X}{Y}}\right)}
\right] .
\end{aligned}
\end{equation}
The parameters in eq.~\ref{eq:viewF} are based on the length of the lamp axis,
$l$, its radius, $r$ and the distance from the lamp, $d$. They are calculated as follows:
\begin{equation}
\begin{aligned}
H_l=\frac{d}{r}, \qquad L_l=\frac{l}{r}, \qquad X=(1+H_l^2)+L_l^2\\
Y=(1-H_l^2)+L_l^2,\qquad  M = \sqrt{\frac{H_l-1}{H_l+1}} .
\end{aligned}
\end{equation}
The total view factor, $F_{tot}$, is actually adopted to evaluate the UV--C irradiance 
field reaching a parcel.
For points located between lamp edges, $F_{tot}$ is given by the superposition of 
two different contributions: 
\begin{equation} 
F_{tot}=F(l_1)+F(l_2) .
\end{equation}
It is worth noting that $l_1$ and $l_2$ are segments deriving from lamp
splitting in correspondence of the parcel position. 
Hence, $F(l_1)$ and $F(l_2)$ are the results of the application of eq.~\ref{eq:viewF}
to the lamp portions.
%
Differently, for parcels located beyond or before the lamp ends, a "ghost" length, $l_g$,
is considered. This length is the the axial distance between parcel and lamp edge.
In this case, the total view factor is calculated subtracting the ghost portion contribution as follow:
\begin{equation}
F_{tot}=F(l+l_g)-F(l_g) .
\end{equation}
Lastly, $F_{tot}$ is used to evaluate the UV--C field intensity on each parcel present in the domain
as a function of its distance from the lamp axis:
\begin{equation}
E_p = \frac{W_l}{2\pi r l}F_{tot}
\end{equation}
in the above equation $W_l$ is the total power of the lamp.\\

%
%
%
%
%

\section{Numerical approximation}
\label{sec:num}
The governing equations solution relies on the OpenFOAM
library. Thus, the unstructured, colocated, cell-centred finite
volume method was adopted for the space discretisation.
An implicit, three levels, second–order scheme was used for the 
time–integration together with the dynamic adjustable time
stepping technique for guaranteeing a local Courant ($\mathrm{Co}$) number
less than a user--defined value ($\mathrm{Co_{max}}$).
The interpolation of convective fluxes is treated by the linear upwind scheme, whereas diffusive 
terms are discretised by a standard second--order central scheme.
Moreover, the pressure velocity--coupling is handled through PISO procedure\cite{Issa198640}.\\
For the linear solvers a preconditioned conjugate gradient method (PCG)
with a diagonal incomplete--Cholesky preconditioner was used to solve
pressure equation.
A preconditioned bi--conjugate gradient method (PBiCG) with the DILU 
preconditioner was adopted instead for the remaining equations. 
In particular, a local accuracy of $10^{-7}$ was established for the pressure,
whereas other linear systems were considered as converged when the residuals 
reached the machine precision.\\
\subsection{Computational grids}
\begin{figure}[htbp]
 \centering
 \includegraphics[width=0.45\textwidth]{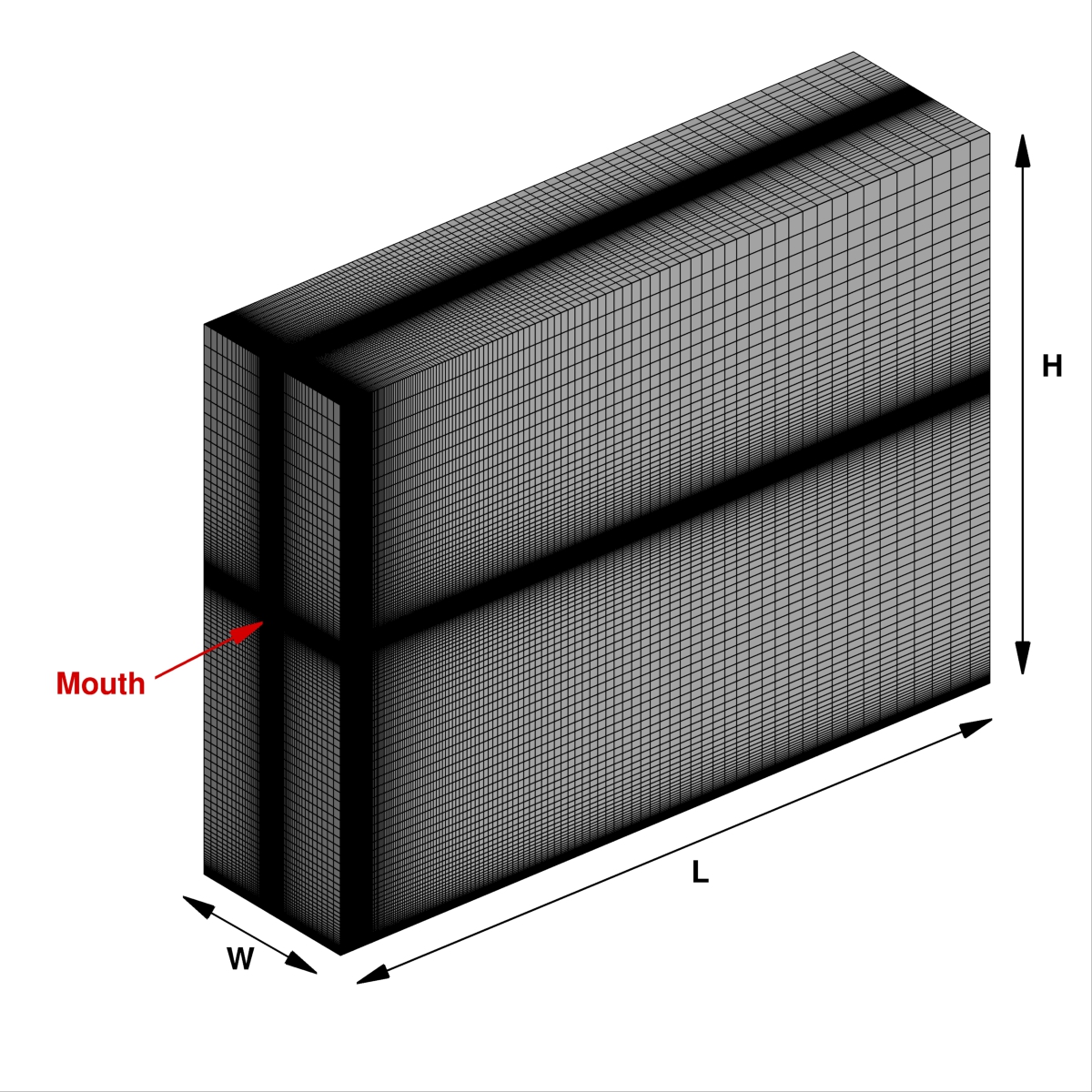}
\caption{Computational grid representation.}
\label{fig:grid1}
\end{figure}

\begin{figure}[htbp]
 \centering
 \includegraphics[width=0.45\textwidth]{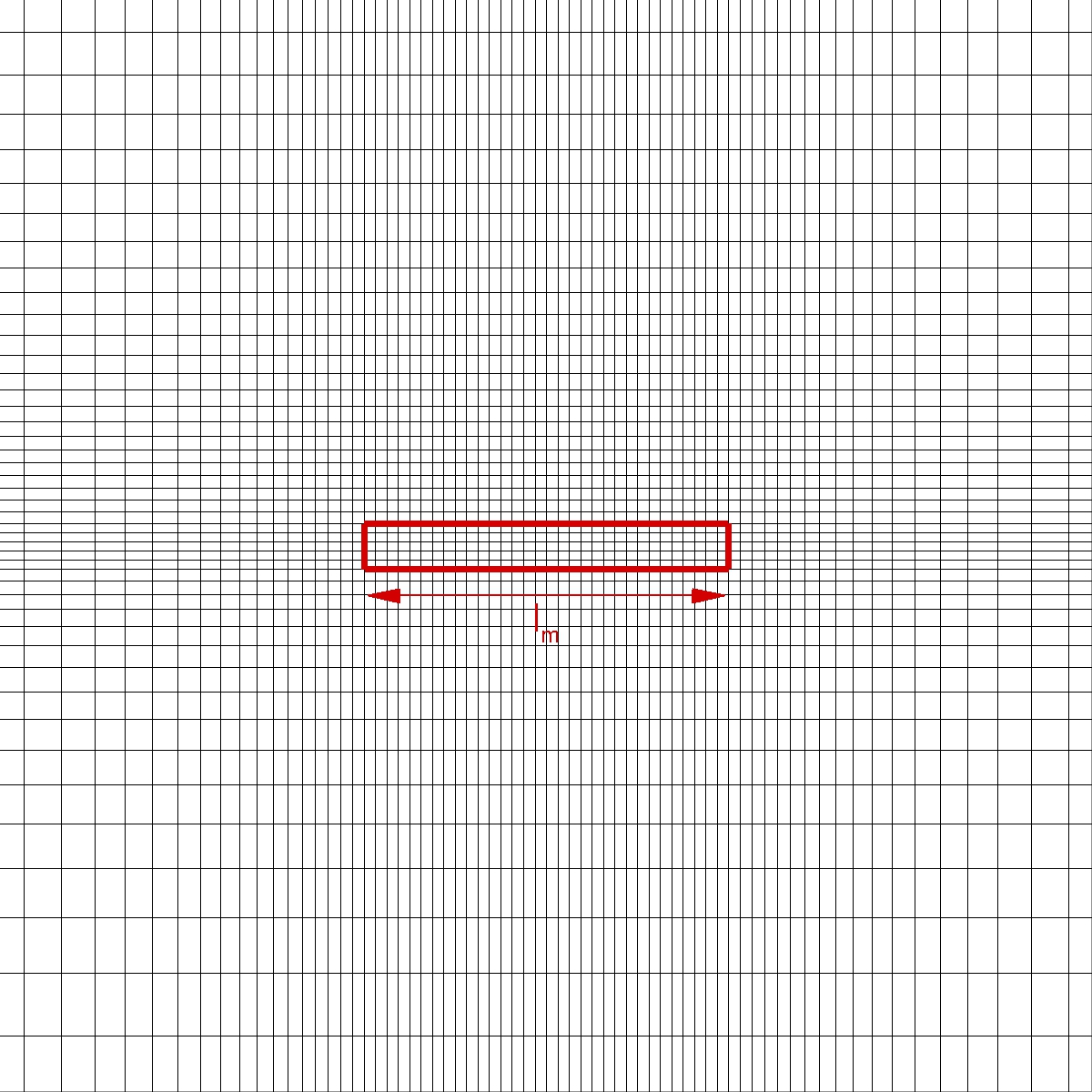}
\caption{Fully-structured mesh, mouth-print refinement.}
\label{fig:grid2}
\end{figure}

In the present work, a 3D computational domain was considered as shown in Fig.~\ref{fig:grid1}.
It consists of an air volume starting from the mouth--print of a standing coughing person.
A length $L=4$~$m$, a width $W=1$~$m$ and a height $H=3$~$m$ were adopted, in accordance with Dbouk and Drikakis~\cite{Dbouk}.
The mouth print is approximated as rectangular, having a length of
$l_m=0.04$~$m$ and a total area $A_m=2\cdot 10^{-4}$~$m^2$.
The reference frame origin, $O=\left(0 , 0 , 0\right)$, is fixed in the same plane where  
the mouth print is placed, see Fig.~\ref{fig:grid1} and Fig.~\ref{fig:grid2}.
In particular, $O$ is in the middle of the previous face in correspondence with the intersection
of the domain bottom side.
$x$--axis is aligned with bio--aerosol propagation direction; $y$--axis represents transverse direction,
while $z$--axis is the vertical direction.\\
The mouth--print centre, $P_m$,  was placed in the same position selected by Dbouk and Drikakis~\cite{Dbouk}: 
 $P_m = (0, 0, 1.63)$.\\
%
Fully--structured meshes were built in order to discretise the domain. 
A suite of three different grids named $S1$, $S2$ and $S3$ was generated 
using, for all the cases, $160$ cells for mouth discretization.
Besides, mesh elements size grading was applied so as to achieve
a proper discretisation in the droplets' emission and transport areas (Fig.~\ref{fig:grid2}).
The height of the first cell next to the ground, $z_c$, was set at $10^{-3}$~$m$.
Other grids details are collected in Tab.~\ref{tab:grid_p}.\\
\begin{table} \caption{Grid point distribution.}
  \label{tab:grid_p}
  \centering
  \begin{tabular}{|c|c|c|c|c|}
     \hline  Grid  & L points & W points & H points & Total \\   
     \hline \hline \hline 
     S1 & 170 & 112 & 150 & $2.856\cdot10^6$ \\
     \hline
     S2 & 221 & 136 & 194 & $5.830\cdot10^6$ \\
     \hline
     S3 & 270 & 158 & 231 & $9.854\cdot10^6$ \\
    \hline
  \end{tabular}
\end{table}

\begin{figure}[htbp]
 \centering
 \includegraphics[width=0.45\textwidth]{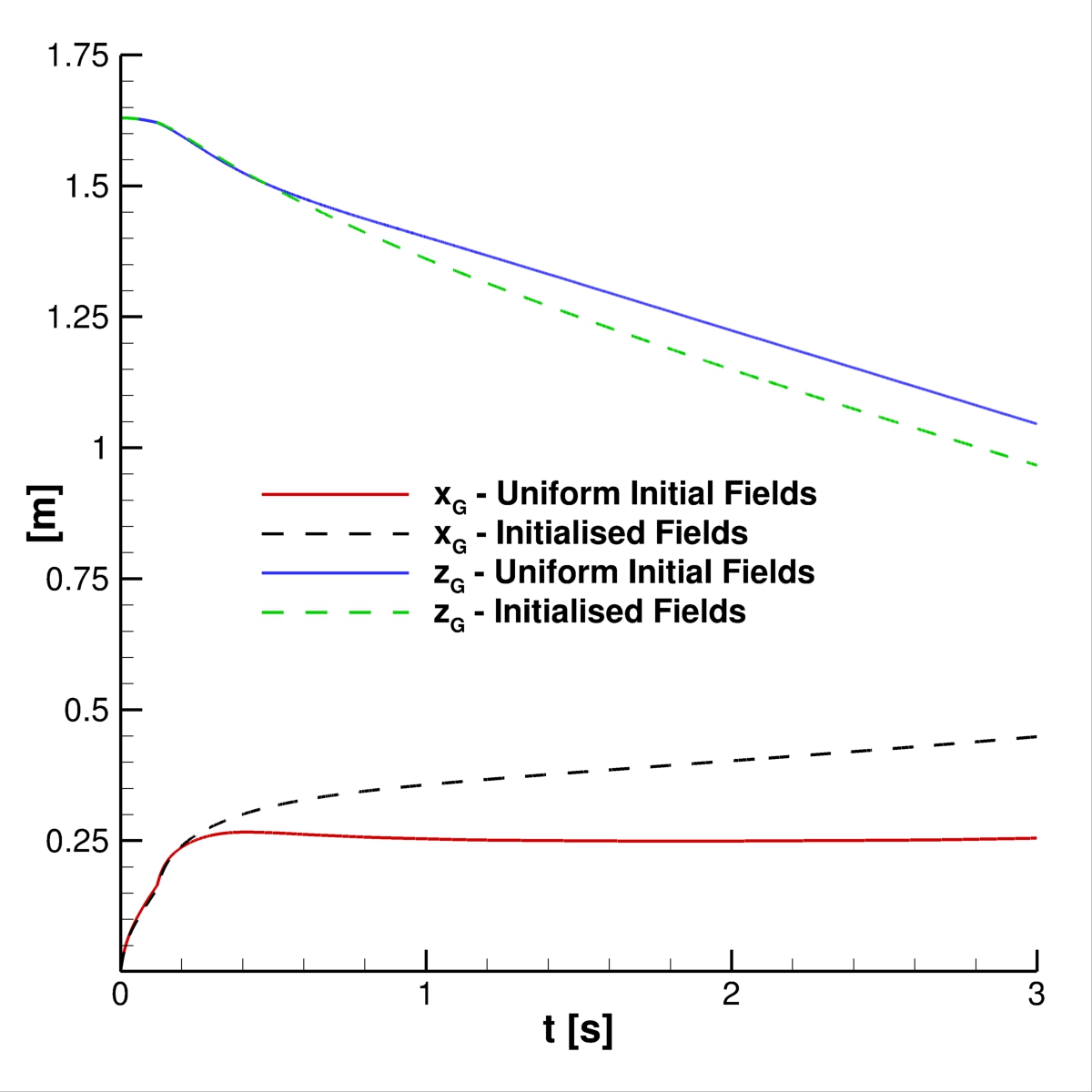}
\caption{Effect of the initial conditions on the particles' cloud evolution. S2 grid, $\mathrm{Co_{max}} = 0.2$.}
\label{fig:precursor}
\end{figure}

\subsection{Initial and boundary conditions}
A stepped velocity inlet at the mouth boundary, with injection of parcels,
was applied to mimic the human cough over $0.12$~$s$.
Velocity inlet value was deduced on the base of measurements carried on by Scharfman et al.\cite{Scharfman2016}, and 
it is equal to $8.5$~$m/s$ in the streamwise direction both for carrier fluid and injected parcels. 
In the same boundary turbulence intensity, $\mathrm{Tu}$, is fixed at $15\%$ and the mixing length equal to $7\cdot 10^{-3}$.
Furthermore, the initial total mass of saliva droplets laden into the domain is $7.7$~$\mathrm{mg}$ 
according to the experimental measurements performed by  Xie et al.\cite{Xie:2009} and CFD simulations of Dbouk and Drikakis\cite{Dbouk}.
Saliva is, in general, a complex fluid but, following Van Der Reijden et al.\cite{VanDerReijden1993141},
it could be approximated as water. 
For this reason, the impact of the UV--C field on the parcels' temperature is neglected 
since UV--C water absorptivity is extremely low. 
The remaining part of the $y$--$z$ plane at $x=0$~$m$ is such that all the variables have 
a null gradient through it.
The bottom side of the domain,~\emph{i.e.} the ground, is modelled as a standard 
wall. Symmetry condition is imposed on lateral boundaries: $y=\pm0.5$~$m$.
On the other hand, zero gradient condition is set for all variables at the domain top with the exception 
of the pressure. In this case, the pressure is reduced of its hydrostatic level.
The $y$--$z$ plane at $x=4$~$m$ is managed as a physical outflow.
However, the pressure is imposed to decrease linearly, starting from atmospheric pressure 
level at $z=0$.\\
The initial temperature of the carrier fluid is $20^\circ$ $\mathrm{C}$ with relative humidity
fixed at $50\%$. The ground is at $25^\circ$ $\mathrm{C}$, while the air and droplets 
ejected by human mouth are at $34^\circ$ $\mathrm{C}$.
The initial mass fraction composition of the Eulerian phase is: $0.991$ dry--air
and $0.009$ water--vapor as in Dbouk and Drikakis~\cite{Dbouk}. 
The hypothesis of injection of saturated moist--air from the mouth is also taken into 
account in the present work.\\
The estimated maximum Weber number is smaller than the critical one\cite{Sula2020}, this 
is the reason why any secondary breakup model is introduced in the following computations.
It is very important to put in evidence that previous initial conditions 
are not adopted for our full computations. 
Indeed, they are used as initial conditions of a preliminary simulation aimed 
to generate adequate initial fields the full runs.
In this simulation, the transient condition
for mouth print boundary is not employed and treated like the remaining
part of the $x=0$  plane.
The preliminary simulations stage is considered 
completed when a physical time of $15$~$\mathrm{s}$ is reached.
It is worth emphasising  that, after this precursor,
turbulent variables are initialised, and the hydrostatic pressure field, 
not available in the previously described situation associated to an imposed uniform field, is obtained.
Cloud evolution is also strongly influenced as represented in Fig.~\ref{fig:precursor}. 
Indeed, the cloud centre of mass position (defined in the next Section) in time completely 
changes when initialised fields are employed.  

\section{Results}\label{sec:results}
The present section shows the obtained numerical results referred to the
bio--aerosol produced during coughing as cloud.
Several cloud characteristics will be calculated and 
considered in order to investigate its diffusion and interaction
with artificial UV--C light,~\emph{i.e.}
\begin{inparaenum}[(i)]
\item cloud centre of mass; 
\item streamwise liquid penetration length;
\item fraction of particles present in a reference volume;
\item active fraction in a reference volume. \\
\end{inparaenum}
The cloud centre of mass is computed as follows:
\begin{equation} \label{eq:cloud_OG} 
{\bf G} = \frac{ \sum_{i=1}^ {\widehat{N}_{p}} m_{P,i} {\bf x}_{P,i} } { \sum_{i=1}^ {\widehat{N}_{p}} m_{P,i}   } , 
\end{equation}
in eq.~\ref{eq:cloud_OG} $\widehat{N}_{p}$ is the overall number of parcels laden in the domain 
in a given time--instant. 
In the following lines ${\bf G} = \left( x_G , y_G , z_G \right)$ is considered as the centre of mass components.\\
Streamwise liquid penetration length, $\mathrm{LPL_x}$, is defined as the maximum distance travelled along $x$--axis
by a parcel conserving at least $95$~$\%$ of its initial mass.
It is interesting to put in evidence that for this parameter, publicly available OpenFOAM functions are not used.
Thus, an inline function leaning on the parcel mass stored at the domain immission is developed.
Two different indexes for describing the saliva droplets' population/activation are introduced.
The first index is the ratio between the number of particles present in a reference volume, $\Omega_i$, and the total 
 number of particles in the overall domain, $\Omega_0$, in a given time instant:
\begin{equation}\label{eq:phi_}
\Phi_{\Omega_i} = \frac{ \sum_{k=1}^{  \widehat{N_p} \left( {\Omega_i} \right) } N_{p,k}  }{  \sum_{k=1}^{  \widehat{N_p} \left( {\Omega_0} \right) } N_{p,k}   }
\end{equation}
a second reference index is expressed in the following equation:
\begin{equation}\label{eq:phi_}
\Phi_{A,ij} = \frac{ \sum_{k=1}^{  \widehat{N_p} \left( {\Omega_j} \right) } N_{a,k}  }{  \sum_{k=1}^{  \widehat{N_p} \left( {\Omega_i} \right) } N_{p,k}   } 
\end{equation}
$\Phi_{A,ij}$ is the ratio of active particles in $\Omega_j$ and the number of particles hosted 
in $\Omega_i$. The aim of $\Phi_{A,ij}$ index is to provide a quantitative analysis 
of the impact of  UV--C related biological inactivation.\\
It is essential to remark that the data presented in this paper are focused on parallelpipedal reference volumes having the following features:
\begin{equation}\label{eq:domain}
{\Omega_i} =  [0 , \alpha_i] \times  \left[-0.5 , 0.5\right]  \times \left[1.3 , 1.8 \right]
\end{equation}
the parameter $\alpha_i$, appearing in eq.~\ref{eq:domain}, spans the following values: $0.5$~$m$, $1.0$~$m$, $1.2$~$m$, $1.5$~$m$ which 
are selected in order to investigate a proper safety distance to be held in context of SARS--CoV--2 transmission
containment.
The transverse direction range is considered in order to completely cover the domain.
Lastly, $z$ axis interval is defined for acting on a sufficiently wide range of possible virus receivers' heights.\\
All the computations were performed on the HPC--system CRESCO6 hosted by ENEA at Portici (Italy). 
CRESCO6 comprises 434 nodes with two Intel Xeon Platinum 8160 24--core processors of the Skylake (SKL) generation operating at 2.1 GHz for each node. 
There are 192 GB of RAM available in standard nodes.
The codes were built using Intel compilers and the MPI library version developed by Intel.

\subsection{Grid convergence study}\label{subsec:grid_conv}
\begin{figure}[htbp]
 \centering
 \includegraphics[width=0.45\textwidth]{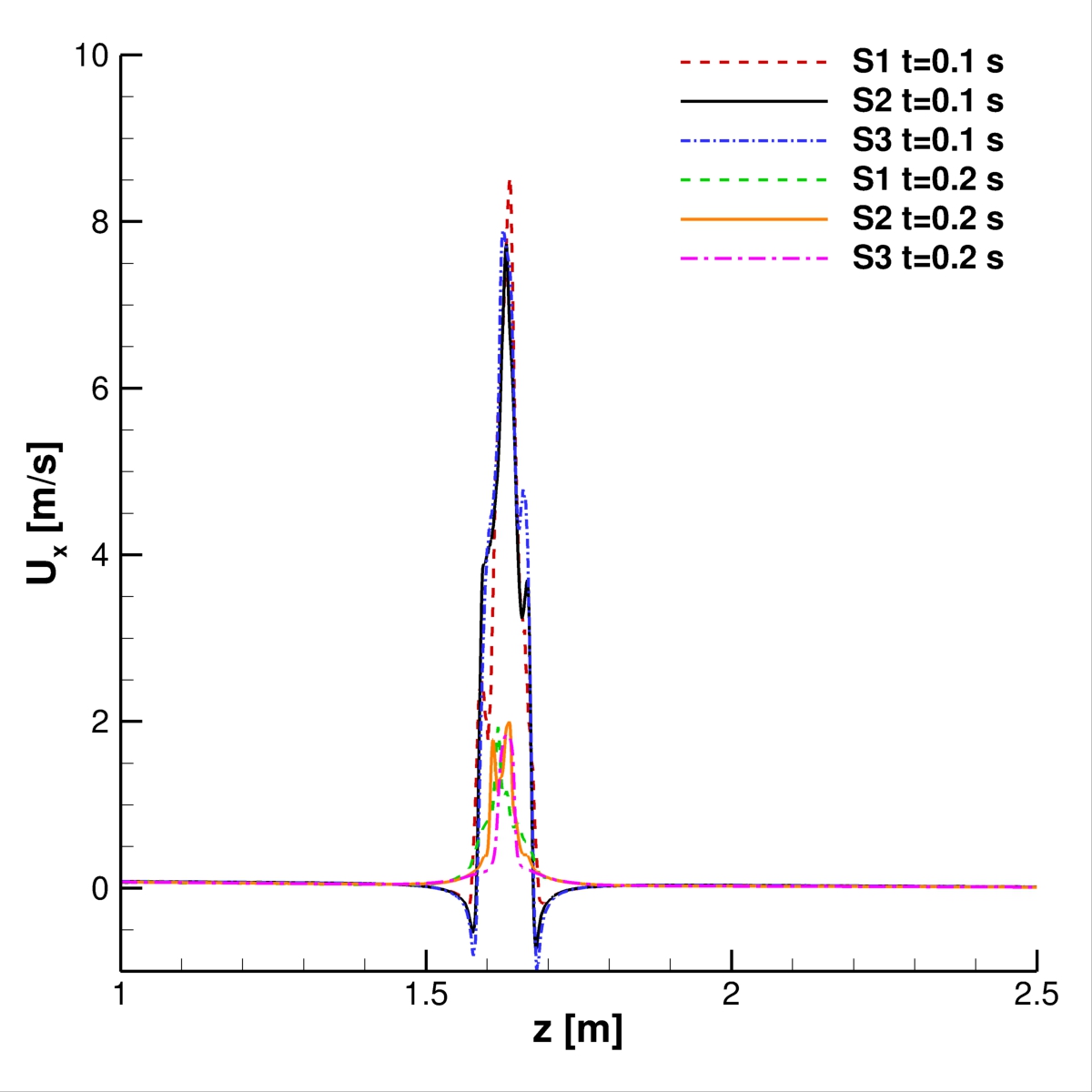}
\caption{Streamwise velocity profiles. $x=0.15$~$\mathrm{m}$, $y=0$~$\mathrm{m}$, $t=0.1$~$\mathrm{s}$, $0.2$~$\mathrm{s}$.}
\label{fig:ux_x=0.15_1}
\end{figure}

\begin{figure}[htbp]
 \centering
 \includegraphics[width=0.45\textwidth]{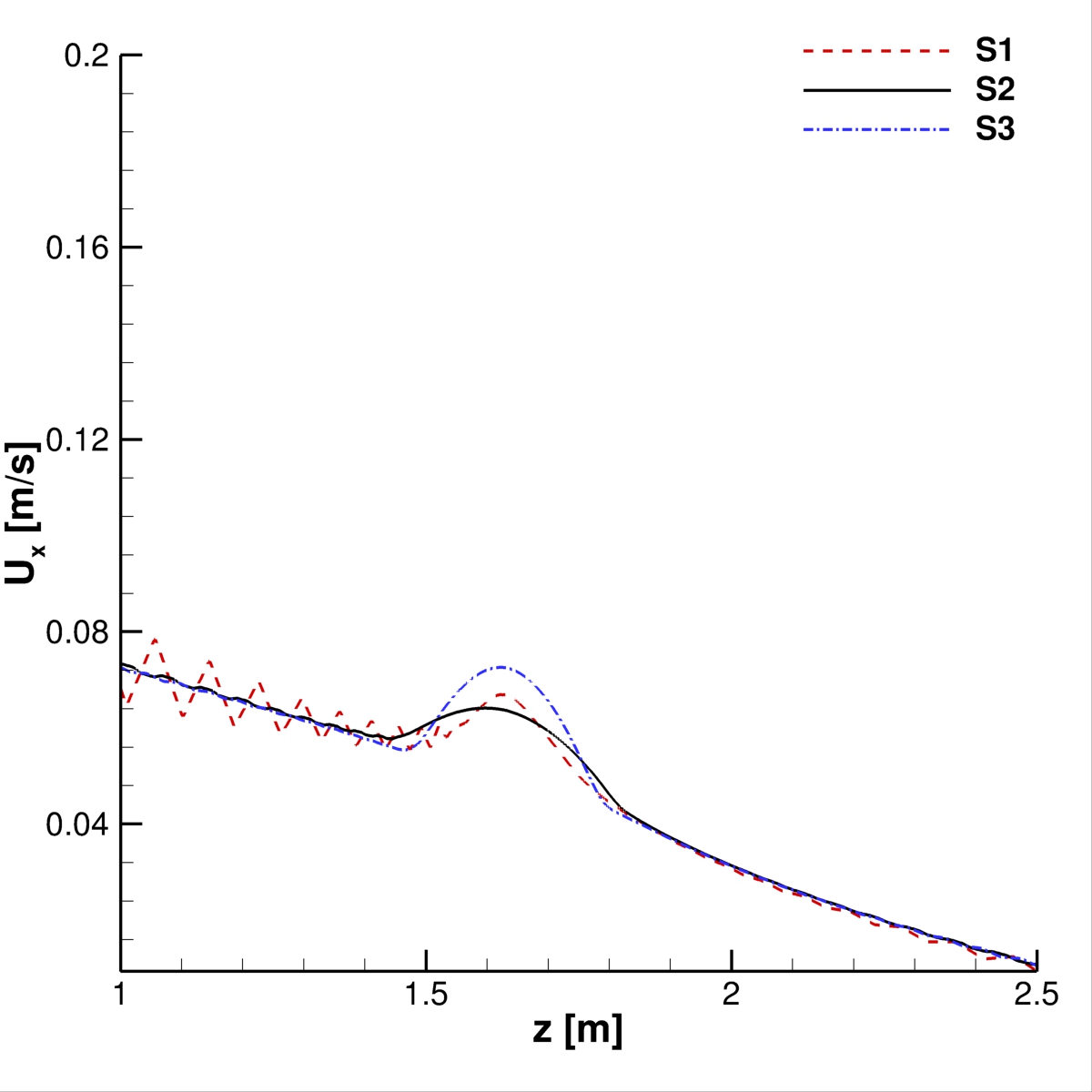}
\caption{Streamwise velocity profiles. $x=0.15$~$\mathrm{m}$, $y=0$~$\mathrm{m}$, $t=1$~$\mathrm{s}$.}
\label{fig:ux_x=0.15_2}
\end{figure}

A grid convergence study was carried out.
Streamwise velocity profiles for the continuous phase were collected in order to  evaluate the time evolution of the cough.
In Fig.~\ref{fig:ux_x=0.15_1} and Fig.~\ref{fig:ux_x=0.15_2} the streamwise velocity, $U_x$, behaviour in the mouth--print proximity is showed:
$S2$ and $S3$ results are in good agreement both during the ejection phase and immediately after it has occurred. 
$S1$ grid, instead, produces unphysical fluctuations in the entertainment zone. 
This is also true when the primary flux evolves into the domain (Fig.~\ref{fig:ux_t1_1} -\ref{fig:ux_t1_2}).

\begin{figure}[htbp]
 \centering
 {\includegraphics[width=0.45\textwidth]{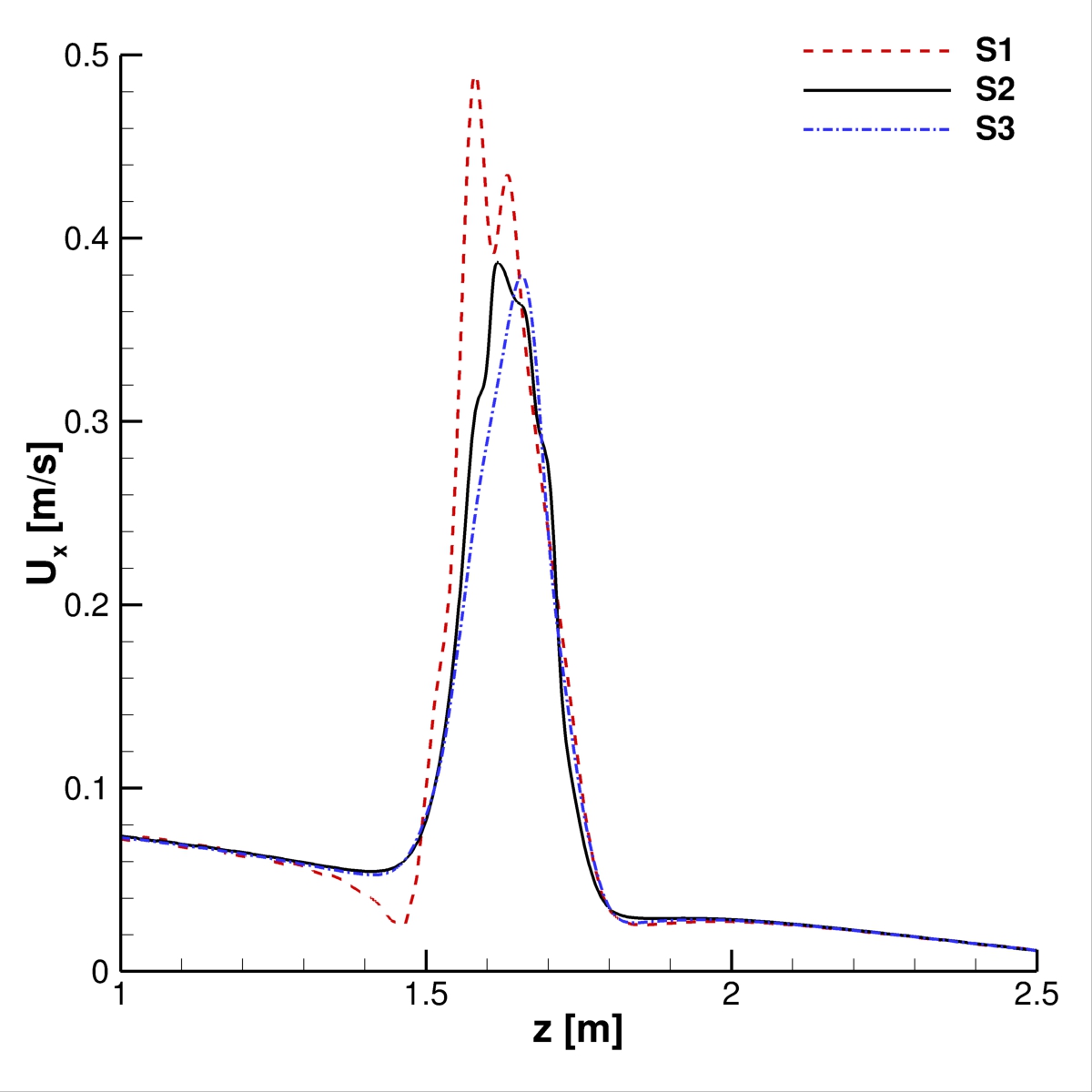}\label{fig:ux_t1_1}}
\caption{Streamwise velocity profiles. $x=0.5$~$\mathrm{m}$, $y=0$~$\mathrm{m}$, $t=1$~$\mathrm{s}$.}
\label{fig:ux_t1_1}
\end{figure}

\begin{figure}[htbp]
 \centering
 {\includegraphics[width=0.45\textwidth]{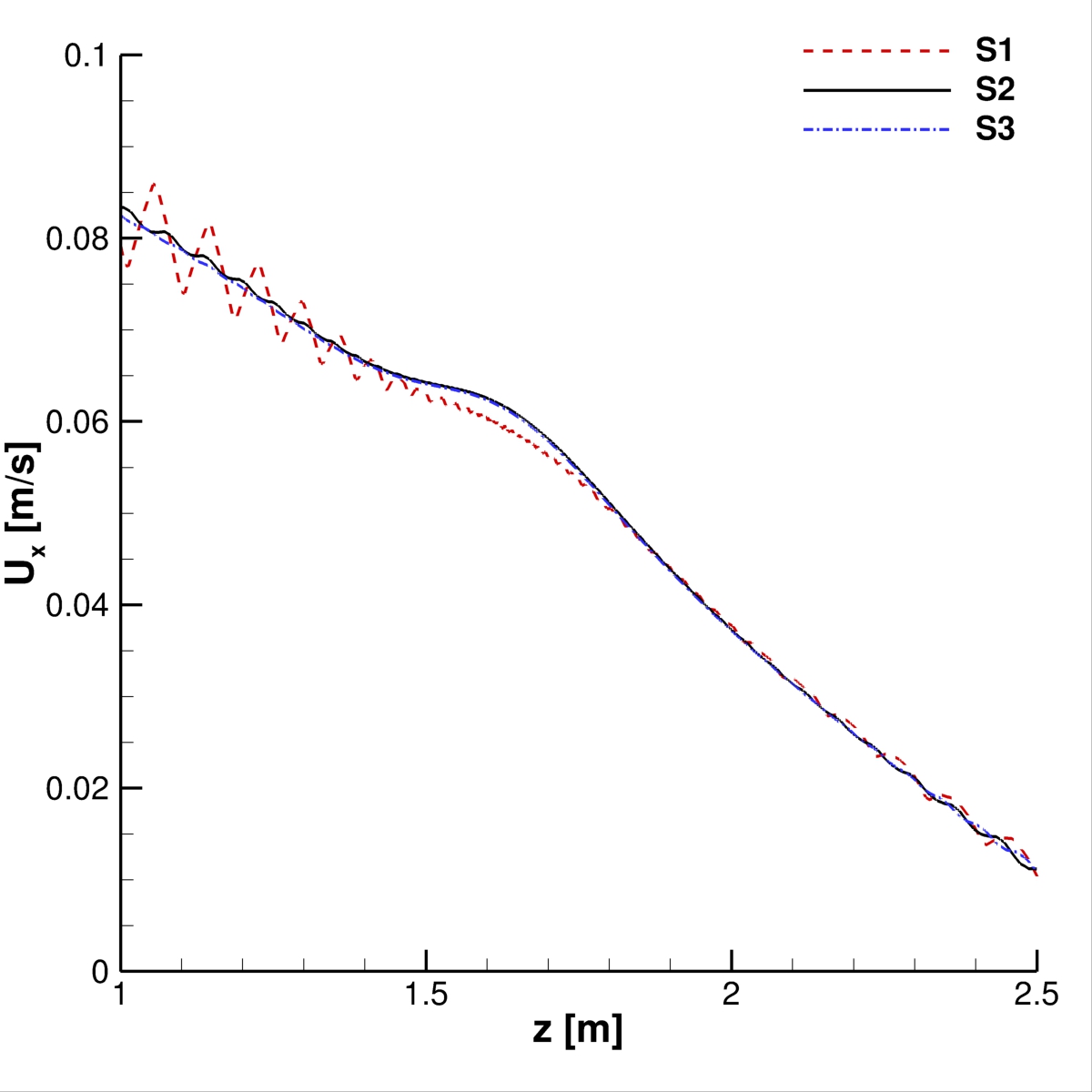}\label{fig:ux_t1_2}}
\caption{Streamwise velocity profiles. $x=1$~$\mathrm{m}$, $y=0$~$\mathrm{m}$, $t=1$~$\mathrm{s}$.}
\label{fig:ux_t1_2}
\end{figure}
For what concerns the cloud development, its centre of mass position and $\mathrm{LPL_x}$ time evolution were monitored. 
Examining Fig.~\ref{fig:grid_cog} is easy to see that $S2$ and $S3$ grids provide very similar behaviour. $S1$, instead,
underestimates  ${\bf G}$, with a major impact on $x_G$.
This is confirmed once again from the comparison in  Fig.~\ref{fig:grid_lpl}. 
For this, $S2$ was chosen as the best compromise between solution accuracy and computational load
for all the following simulations.\\
It is important to point out that, for the entire grid convergence study, 
$\mathrm{Co_{max}} = 0.4$  was adopted for computational efficiency reasons.
\begin{figure}[htbp]
 \centering
 {\includegraphics[width=0.45\textwidth]{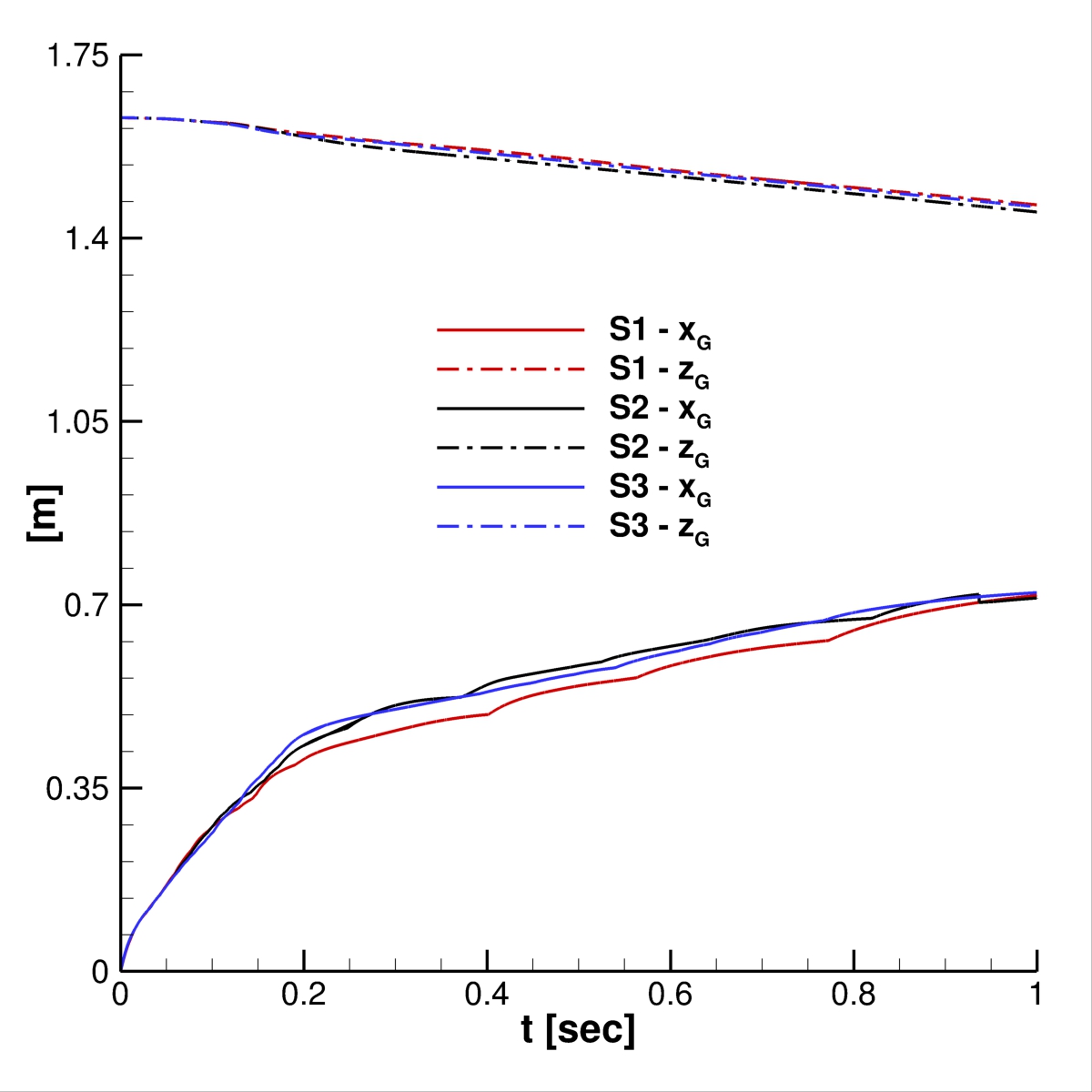}\label{fig:grid_cog}}
\caption{Grid effect. Particles' cloud properties: centre of mass positions.}
\label{fig:grid_cog}
\end{figure}

\begin{figure}[htbp]
 \centering
 {\includegraphics[width=0.45\textwidth]{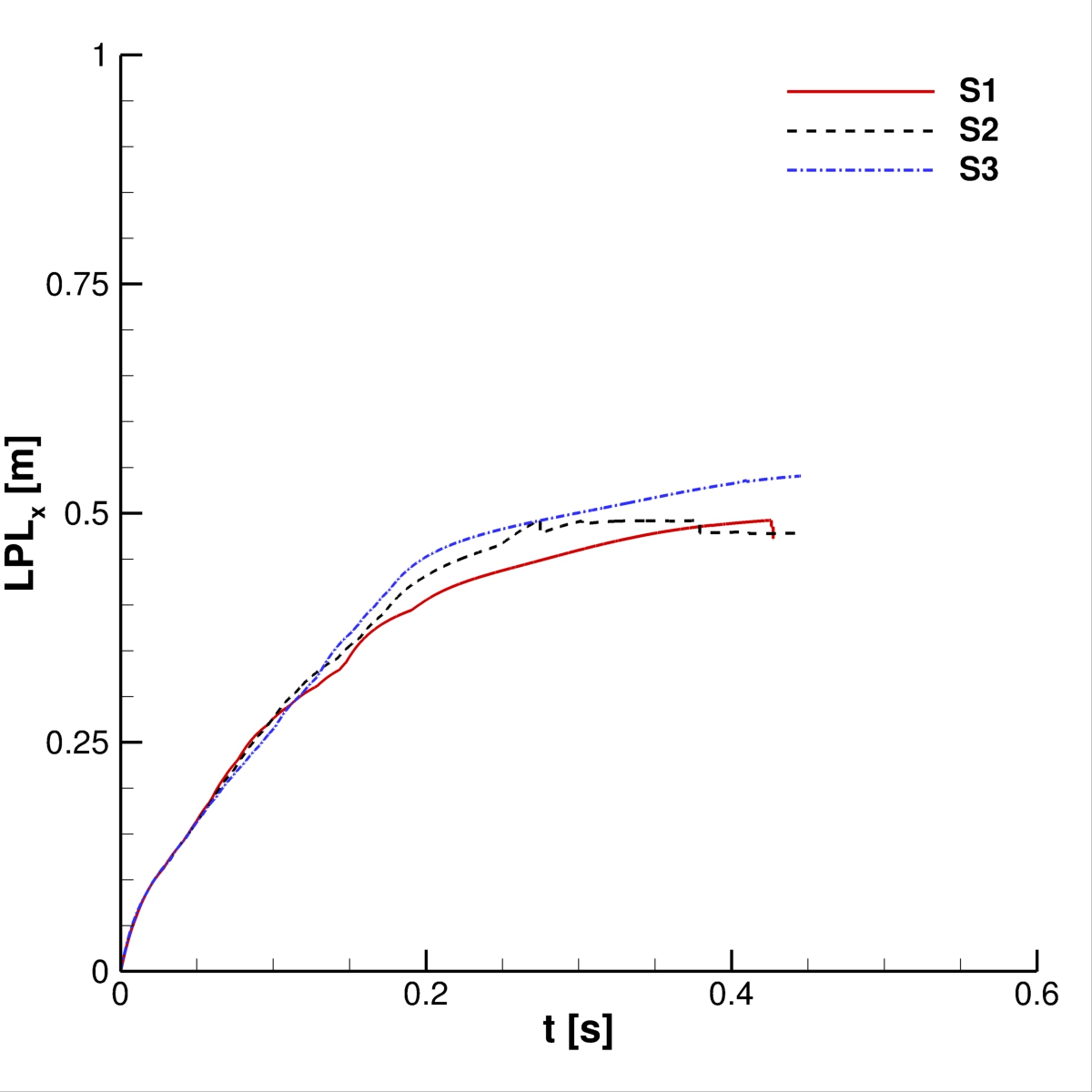}\label{fig:grid_lpl}}
\caption{Grid effect. Particles' cloud properties: streamwise liquid penetration length.}
\label{fig:grid_lpl}
\end{figure}
\subsection{Courant number effect}\label{subsec:Co_conv}
\begin{figure}[htbp]
 \centering
 {\includegraphics[width=0.45\textwidth]{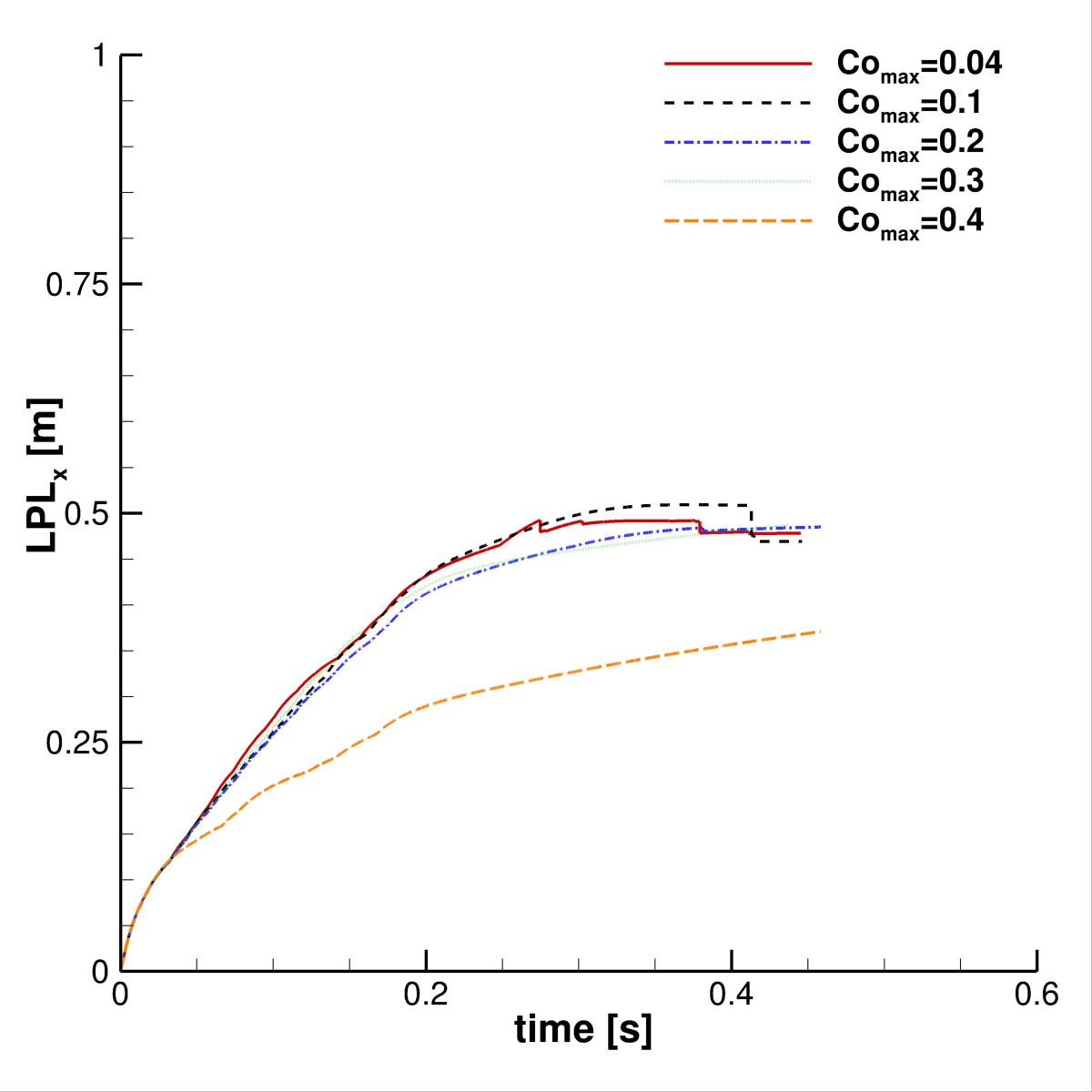}}
\caption{$\mathrm{Co}$ number effect. Particles' cloud properties: streamwise liquid penetration length.}
\label{fig:co_LPL}
\end{figure}

In the context of the dynamic adjustable time stepping technique adopted in this work,
the evaluation of the correct $\mathrm{Co_{max}}$ is crucial. 
Five different maximum Courant numbers: $0.04,\;0.1,\;0.2,\;0.3,\;0.4$ were evaluated 
and their influence on the cloud properties was analysed.
Fig.~\ref{fig:co_cog} put in evidence that, for $\mathrm{Co_{max}} \leqslant 0.2$, ${\bf G}$ shows the same trajectory.
In fact $x_G$ as well as $z_G$ time evolution are almost indistinguishable for the first three cases.
A slightly different trend was found in the streamwise liquid penetration length (Fig.~\ref{fig:co_LPL}).
In this case, $\mathrm{LPL_x}$, provided by assuming $\mathrm{Co_{max}} = 0.2$, does not perfectly 
replicates the behaviour given using lower ones. 
However, considering that the $\mathrm{LPL_x}$ behaviour is slightly affected and the overall good results, 
$\mathrm{Co_{max}} = 0.2$ is chosen for the investigations, even for ensure acceptable computing times. 
\begin{figure}[htbp]
 \centering
 {\includegraphics[width=0.45\textwidth]{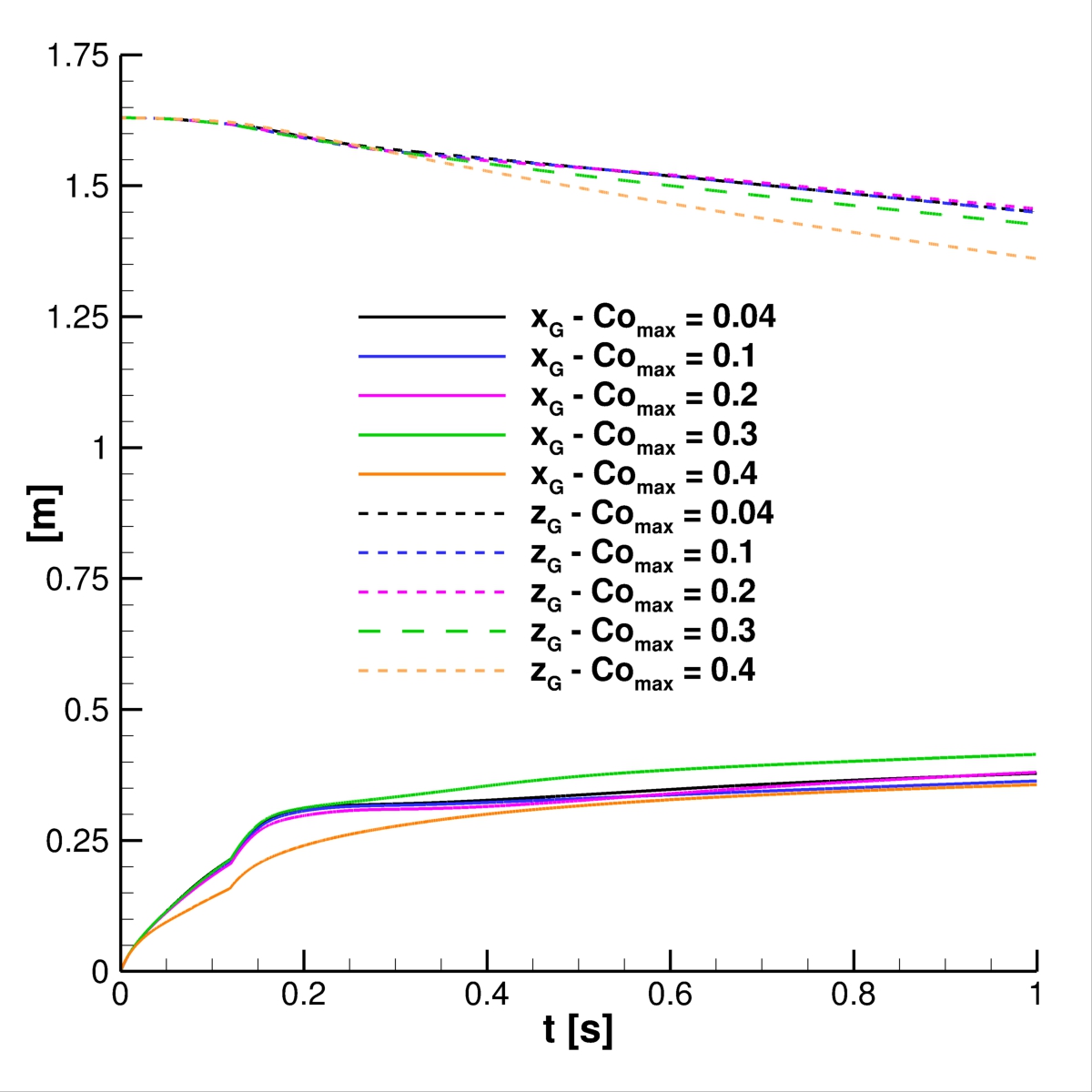}\label{fig:co_cog}}
\caption{$\mathrm{Co}$ number effect. Particles' cloud properties: centre of mass positions.}
\label{fig:co_cog}
\end{figure}

\subsection{Particles per parcel effect}

\begin{figure}[htbp]
 \centering
 {\includegraphics[width=0.45\textwidth]{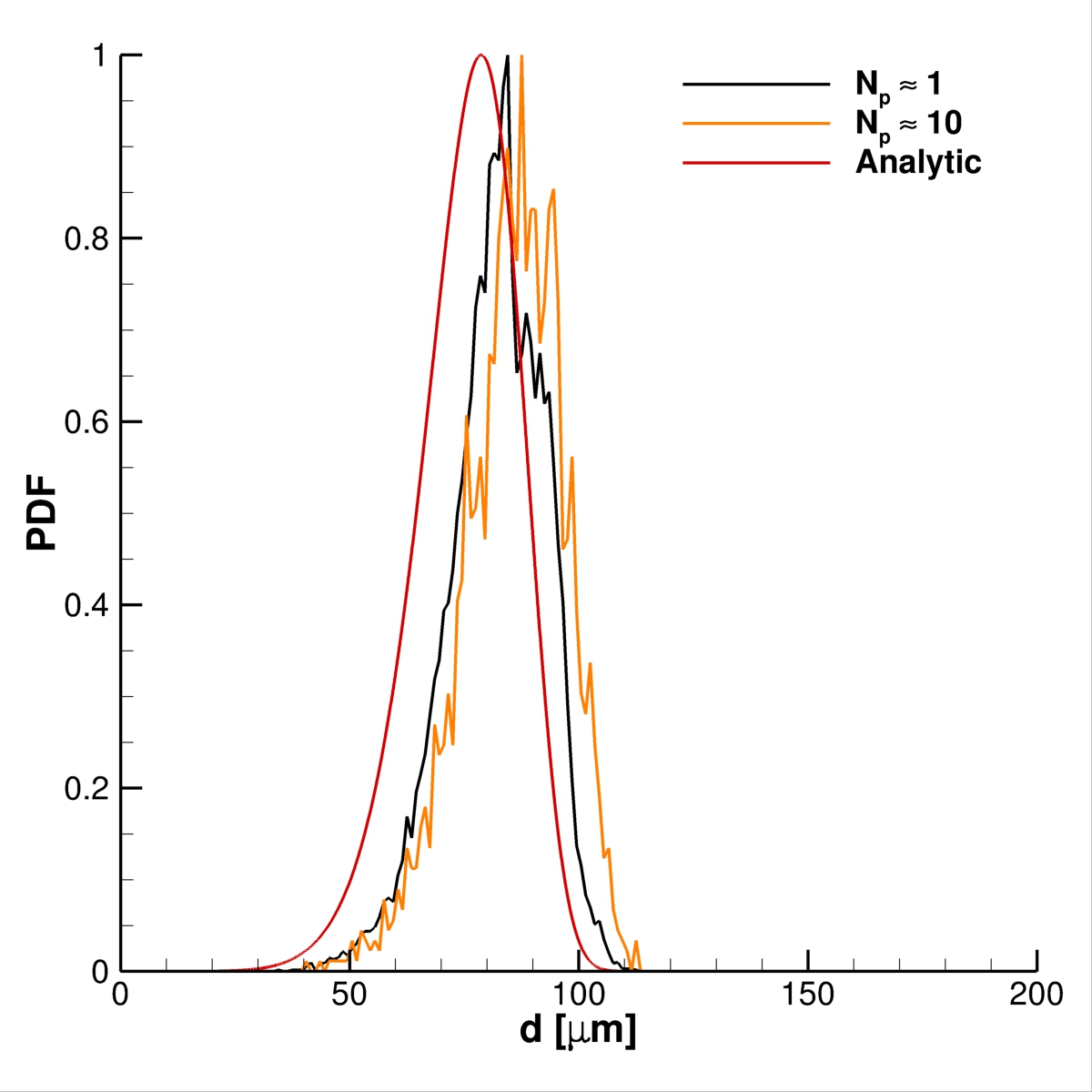}}
\caption{Effect of the number of particles per parcels on diameters PDF.}
\label{fig:ros}
\end{figure}

In the above discussed analyses a mean number of particle per parcel $\overline{N}_{p,i} \approx 10$ 
was considered for the sake of efficiency in the Eulerian--Lagrangian framework coupling.
The intent is to complete the simulation parameters setting by means of the evaluation of $\overline{N}_{p,i}$ 
effect on the droplets' cloud development. It is obvious that for $\overline{N}_{p,i}\rightarrow 1$ each 
particle is independent in its interaction with the carrier fluid and, consequently, the cloud parameters are not affected
by approximations due to the PSI--Cell method parcels based implementation. 
Moreover, as noticeable in Fig.~\ref{fig:ros}, using $\overline{N}_{p,i} \approx 1$, the diameters of the particles
ejected during coughing, better fit the analytic Rosin--Rammler probability density function (PDF) distribution.\\
Finally, $\overline{N}_{p,i} \approx 1$ is adopted for the purpose of achieve the cloud proper modelling in terms of its
dynamical behaviour and PDF diameters size.  

\subsection{Bio--aerosol transport and interaction with UV--C light}
\begin{figure}[htbp]
 \centering
 {\includegraphics[width=0.45\textwidth]{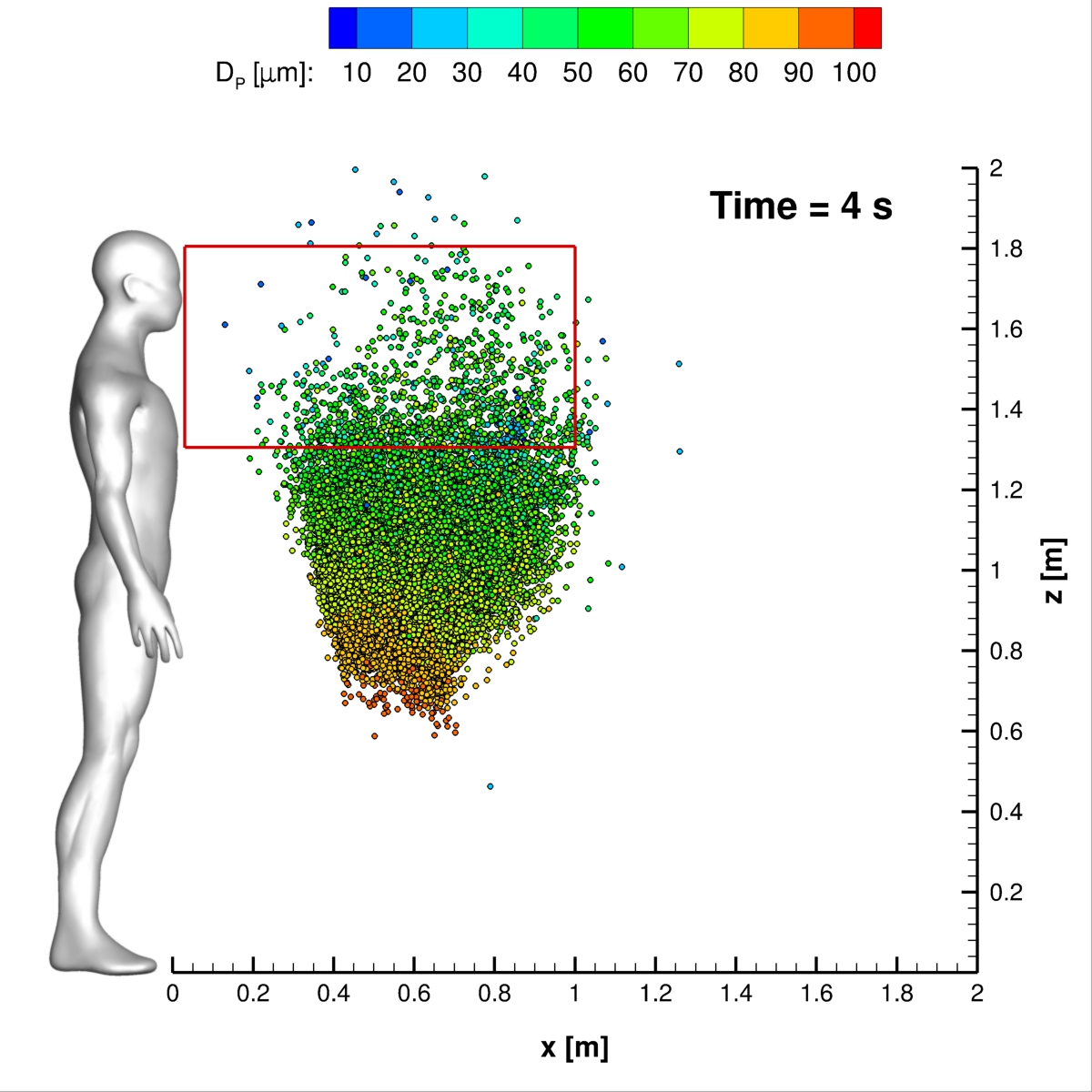}\label{fig:co_lpl}}
\caption{Cloud representation at $t=4$ $s$. Parcels are coloured with particles diameter. Red rectangle is ${\Omega_2}$ volume footprint.}
\label{fig:cloud_d}
\end{figure}

\begin{figure}[htbp]
 \centering
 {\includegraphics[width=0.45\textwidth]{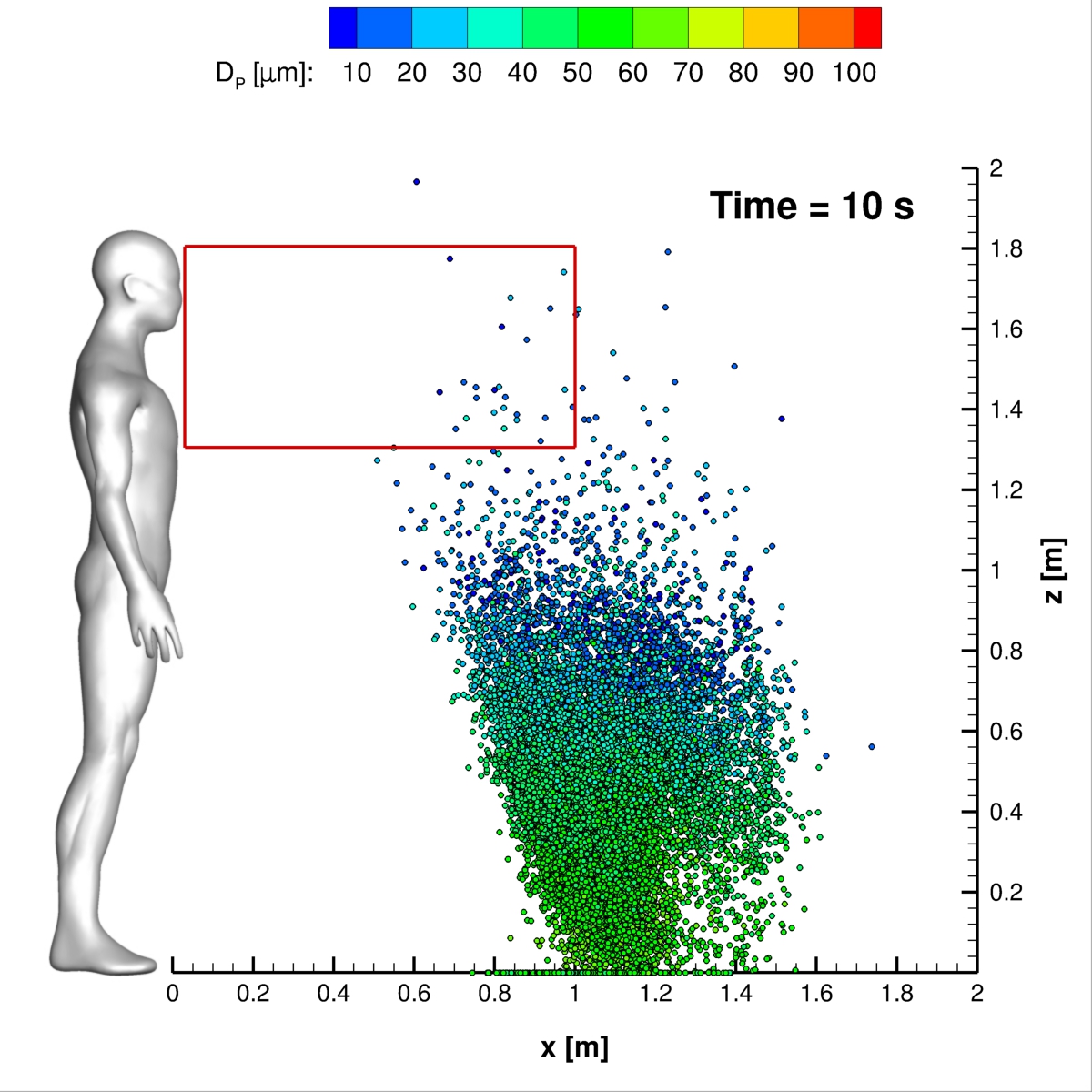}\label{fig:co_lpl}}
\caption{Cloud representation at $t=10$ $s$. Parcels are coloured with particles diameter. Red rectangle is ${\Omega_2}$ volume footprint.}
\label{fig:cloud_d2}
\end{figure}

In this subsection, all relevant features, concerning SARS--CoV--2 transmission of saliva cloud derived
from coughing, are examined. A representation of the computed 
cloud is showed in Fig~\ref{fig:cloud_d}--\ref{fig:cloud_d2}. 

\begin{figure}[htbp]
 \centering
 {\includegraphics[width=0.45\textwidth]{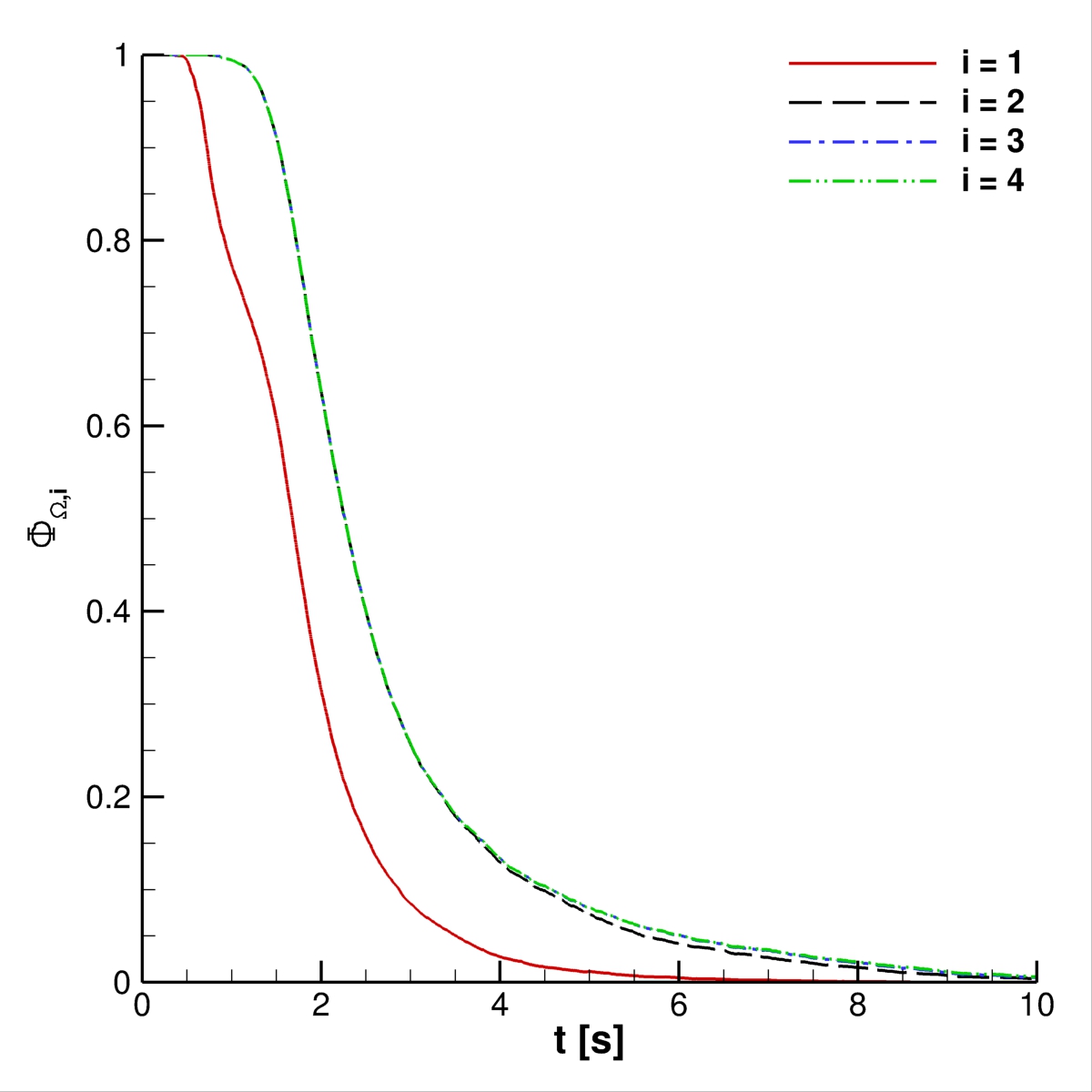}\label{fig:co_lpl}}
\caption{$\Phi_{\Omega_i}$ for one cough ejection.}
\label{fig:fv_omi}
\end{figure}

\begin{figure}[htbp]
 \centering
 {\includegraphics[width=0.45\textwidth]{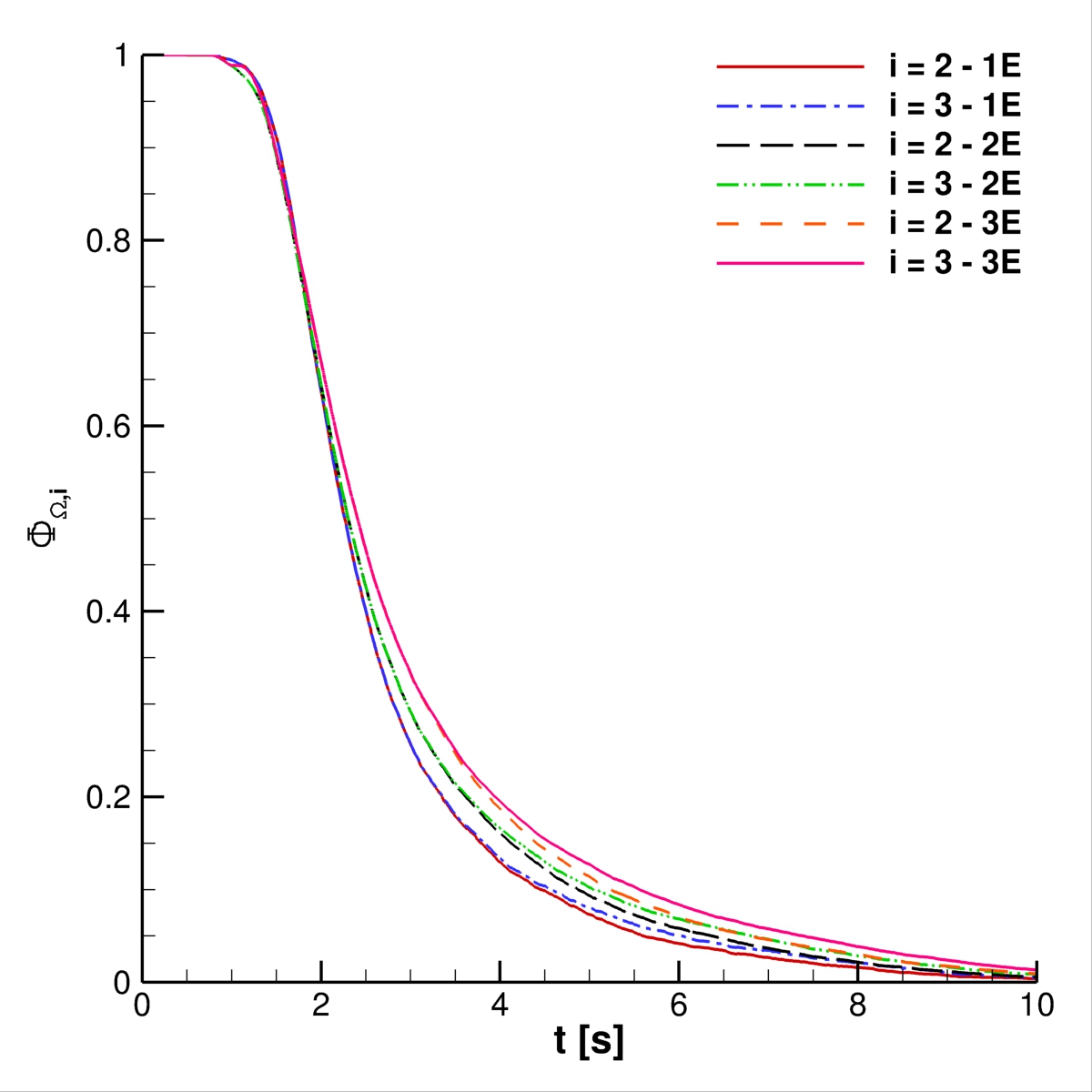}\label{fig:co_lpl}}
\caption{$\Phi_{\Omega_i}$ for multiple cough ejection. In legend $1E$ is for one ejection, $2E$ is for two ejections, $3E$ is for three ejections.}
\label{fig:fv_omi2}
\end{figure}

\begin{figure}[htbp]
 \centering
 {\includegraphics[width=0.45\textwidth]{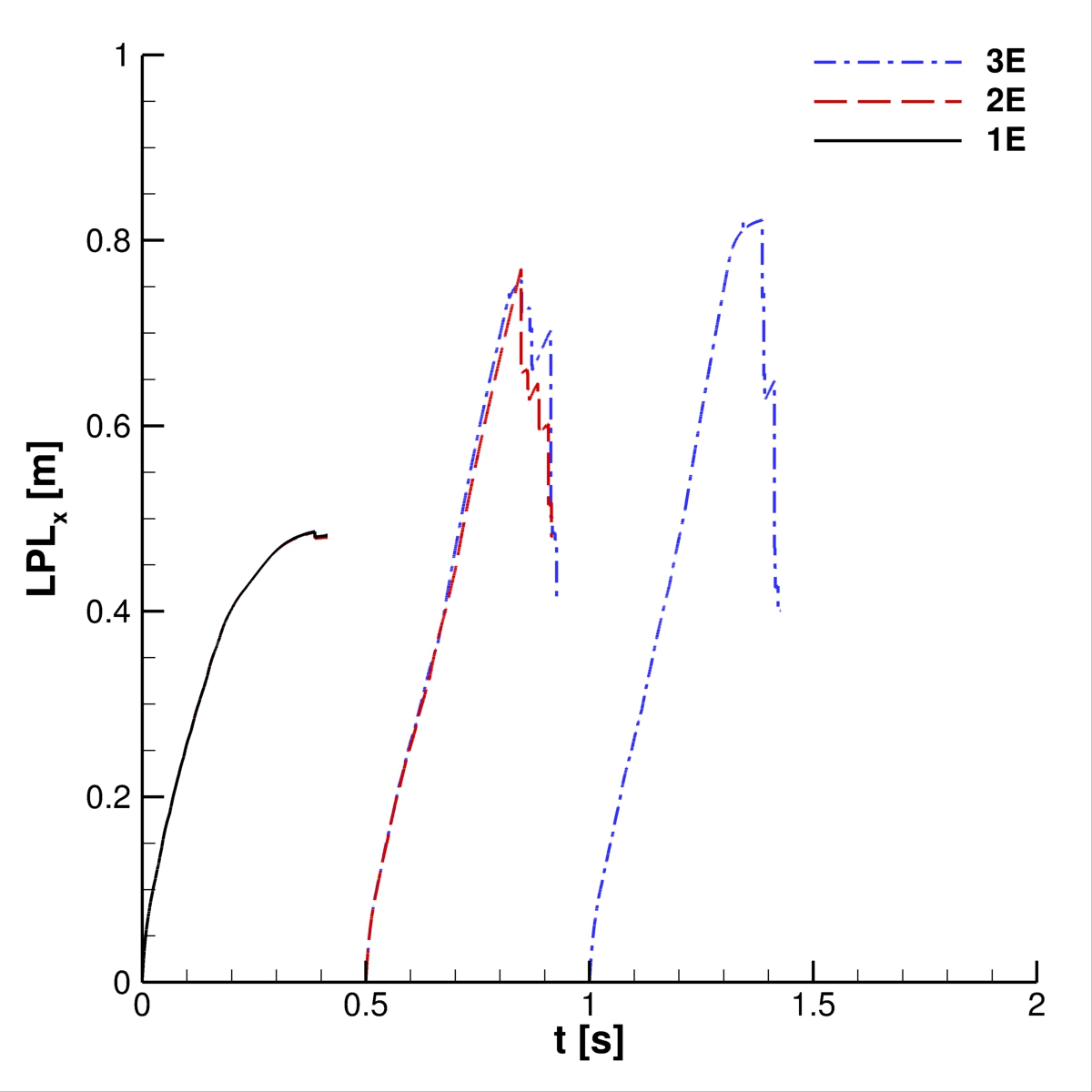}\label{fig:co_lpl}}
\caption{Streamwise liquid penetration length. For legend details see the Fig.~\ref{fig:fv_omi2} caption.}
\label{fig:lpl}
\end{figure}

\begin{figure}[htbp]
 \centering
 {\includegraphics[width=0.45\textwidth]{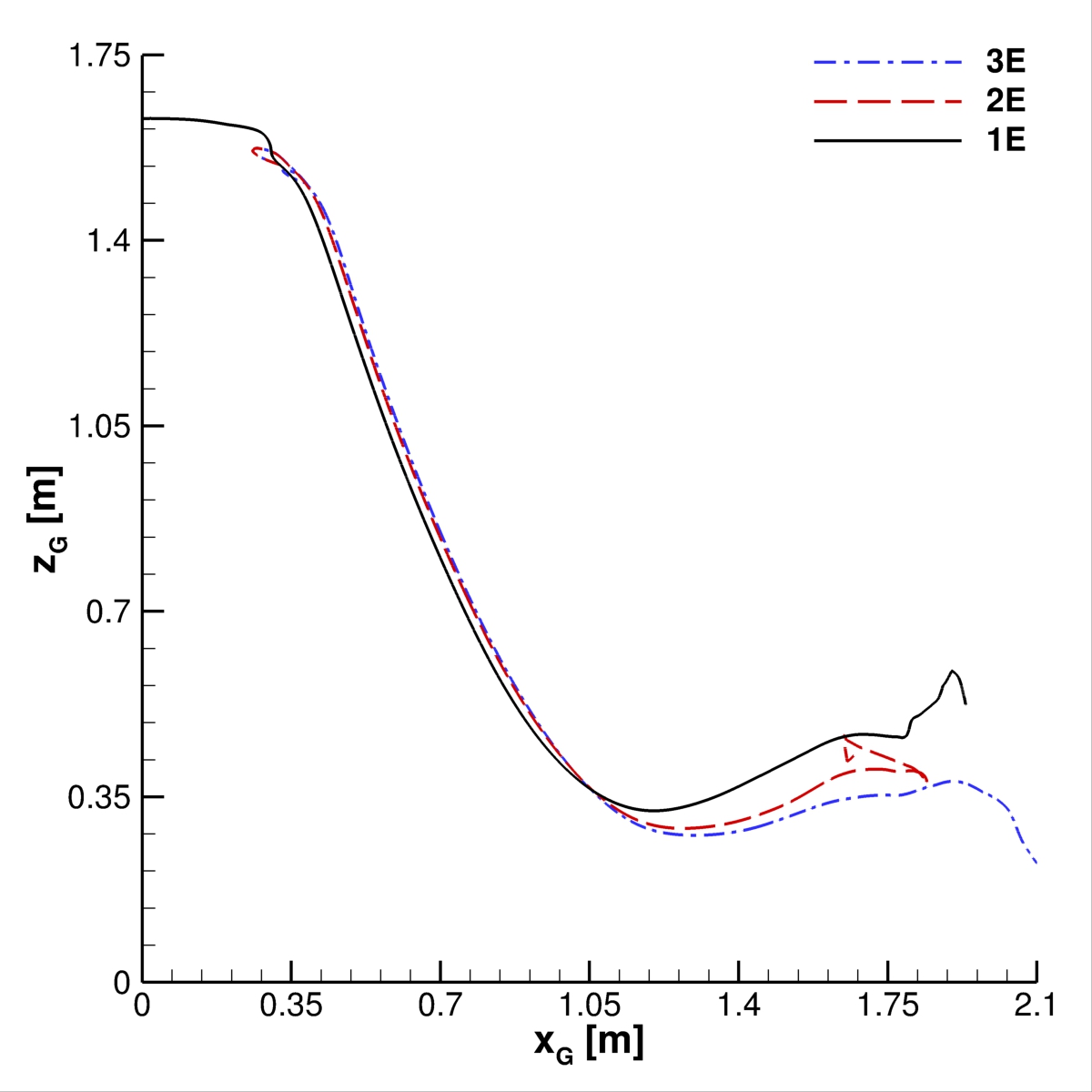}\label{fig:co_lpl}}
\caption{Cloud centre of mass trajectory. For legend details see the Fig.~\ref{fig:fv_omi2} caption.}
\label{fig:og}
\end{figure}

\begin{figure}[htbp]
 \centering
 {\includegraphics[width=0.45\textwidth]{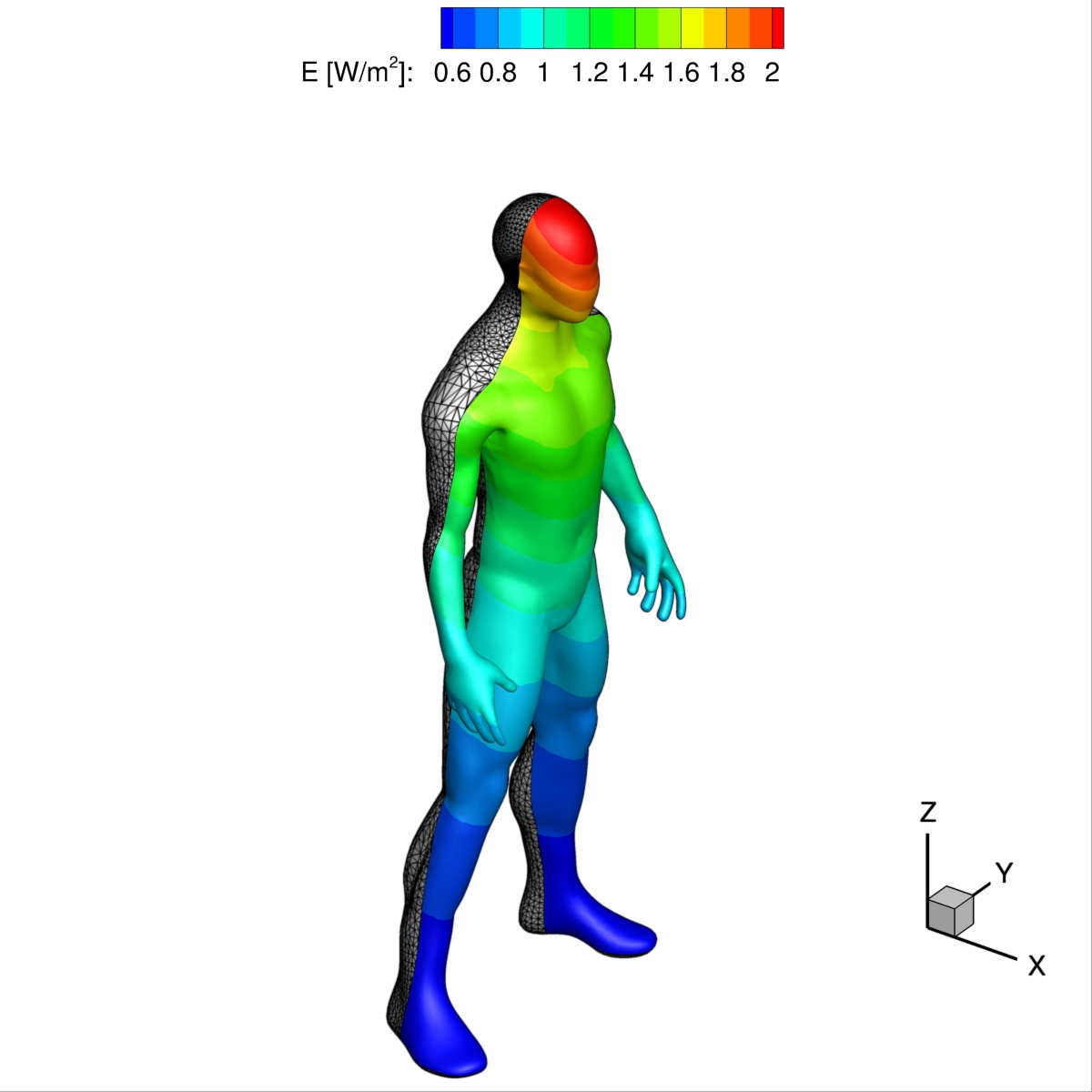}\label{fig:co_lpl}}
\caption{UV--C intensity field received by mannequin placed at domain inlet region. Crossflow oriented $55$ $\mathrm{W}$ lamp.}
\label{fig:cristiano}
\end{figure}

Note that the analyses will not be limited to a single cough ejection but also to multiple
ones.
Up to three cough cycles delayed by $0.38$~$\mathrm{s}$ one over the other\cite{Dbouk1} are applied.
All the presented results are obtained from numerical simulations based on $S2$ grid and
$\mathrm{Co_{max}} = 0.2$ as showed in Secs.~\ref{subsec:grid_conv}--\ref{subsec:Co_conv}. The simulated physical time,
excluding the precursor configuration, is $\sim$$18$ $\mathrm{s}$ (depending on breathing) due to the fact the 
run is stopped when no parcels are still in the domain. 
The total computation time of single case is about $20$ hours run in parallel using $384$ CPU--cores
on CRESCO6.\\
Fig.~\ref{fig:fv_omi} depicts fraction of particles present in the four reference volumes
defined at the beginning of this section.
It is really interesting to note that $\Phi_{\Omega_i}$ curves are very close for $2 \le i \le 4$ . 
So, regarding the possibility to receive infected particles when the external wind is not present,
distances ranging from $1.0$ $\mathrm{m}$ to $1.5$ $\mathrm{m}$ are equivalent.
A similar condition is highlighted in case of multiple cough ejections. Fig.~\ref{fig:fv_omi2}
shows that similar droplets' contamination is produced for ${\Omega_2}$ and ${\Omega_3}$  
for different coughing; ${\Omega_4}$ results are not included in the plot 
to avoid the unreadability of the chart.\\
In Fig.~\ref{fig:lpl} streamwise liquid penetration length, $\mathrm{LPL_x}$, is depicted.
In the case of a single saliva droplets' ejection $O \left(\mathrm{LPL_x} \right) \simeq 0.5$ $\mathrm{m}$ is found;
differently, when the injections' number increases $O \left(\mathrm{LPL_x} \right)$ rises to around $0.8$ $\mathrm{m}$.
It is also interesting to note that, after the first cough, the $\mathrm{LPL_x}$ curve slope grows significantly and 
the largest emitted particles travel not more than $1$$\mathrm{m}$.\\
The centre of mass trajectory on the $x$--$z$ plane is showed in Fig.~\ref{fig:og}. 
The number of cough ejections does not affect largely $x_G, z_G$ evolution except for
its last portion.
For $x_G \simeq 0.35$$\mathrm{m}$, the  $x_G, z_G$ curve underlines a clear trend change. 
Actually, in the proximity of this critical point the trajectory changes from
to almost linear one up to $x_G \simeq 1$$\mathrm{m}$.
This is due to the cancellation of the inertial term.
In the last part of the curve, the evaporation and parcels' interaction with the bottom wall lead to a complex
behaviour. The centre of mass trajectory suggests that several particles are located at distance over $1.5$~$\mathrm{m}$
from the emitter. Nevertheless, a very limited number is situated within the range $1.3\mathrm{m} \le z \le 1.8 \mathrm{m}$.\\
The biological inactivation produced by UV--C light is analysed considering all the saliva
droplets introduced in the domain as fully active. 
A commercial cylindrical lamp (having a radius of $1.4$ $\mathrm{cm}$ and length of $90$ $\mathrm{cm}$) is 
used a UV--C source. It is positioned at the domain top and
different lamp layout were investigated. In the first case, the streamwise orientation was fixed   
and the closer lamp edge to the mouth print is positioned in the point $\left(0.5 \mathrm{m}, 0 \mathrm{m} , 3 \mathrm{m} \right)$.
In the second configuration, the lamp is aligned to the crossflow direction and its centre is placed at $\left( 1 \mathrm{m} , 0 \mathrm{m} , 3 \mathrm{m} \right)$.
Also two different power sizes were considered: $25$ $\mathrm{W}$ and $55$ $\mathrm{W}$.
These values were selected in order to give a limited UV--C dose to a possible cough emitter in 
accordance with Ultraviolet Radiation Guide published by Navy Environmental Health Center (USA)\cite{UVCguide}.
Looking at Tab.~\ref{tab:dose} is possible to observe the average and maximum UV--C dose (received in $10$ $\mathrm{s}$) related 
to mannequin placed at domain inlet region.
Table entries, connected to the average values, are deduced from a numerical integration of UV--C field irradiating the 
space discretisated mannequin (only in its anterior portion) represented in Fig.~\ref{fig:cristiano}.\\
\begin{table} \caption{Average and maximum UV--C dose (in $10$ $\mathrm{s}$) for the
mannequin represented in Fig.~\ref{fig:cristiano}.}
  \label{tab:dose}
  \centering
  \begin{tabular}{|l|c|c|}
     \hline  Case  & Avg. dose $[J/m^2]$ & Max. dose $[J/m^2]$\\
     \hline \hline \hline
     Streamwise - 55W & $8.37$ & $15.21$\\
     \hline
     Crossflow - 55W & $10.16$ & $20.89$\\
     \hline
     Crossflow - 25W & $4.62$ & $9.49$\\
    \hline
  \end{tabular}
\end{table}

\begin{figure}[htbp]
 \centering
 {\includegraphics[width=0.45\textwidth]{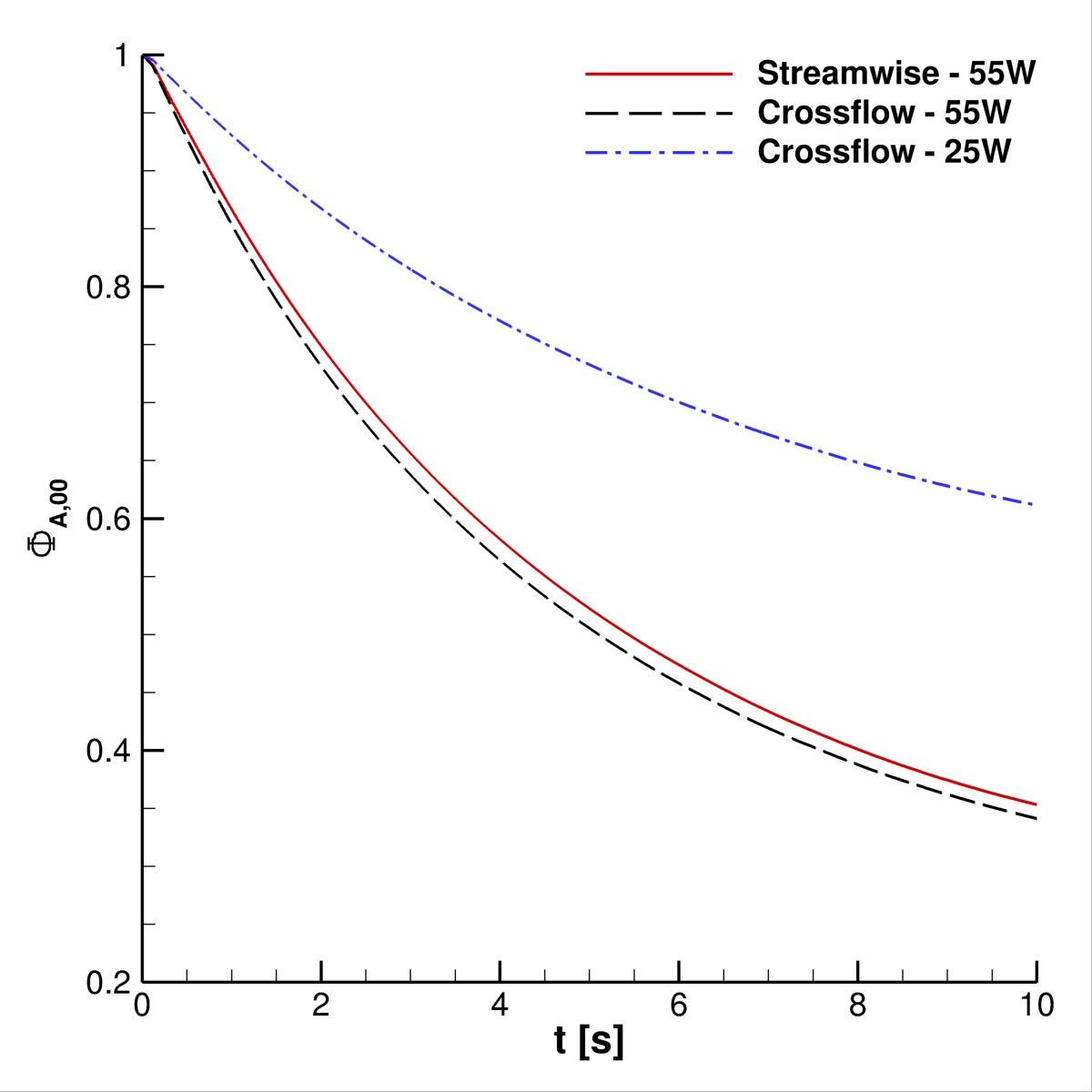}\label{fig:co_lpl}}
\caption{Effect of UV--C lamp power.}
\label{fig:uvc_pow}
\end{figure}

\begin{figure}[htbp]
 \centering
 {\includegraphics[width=0.45\textwidth]{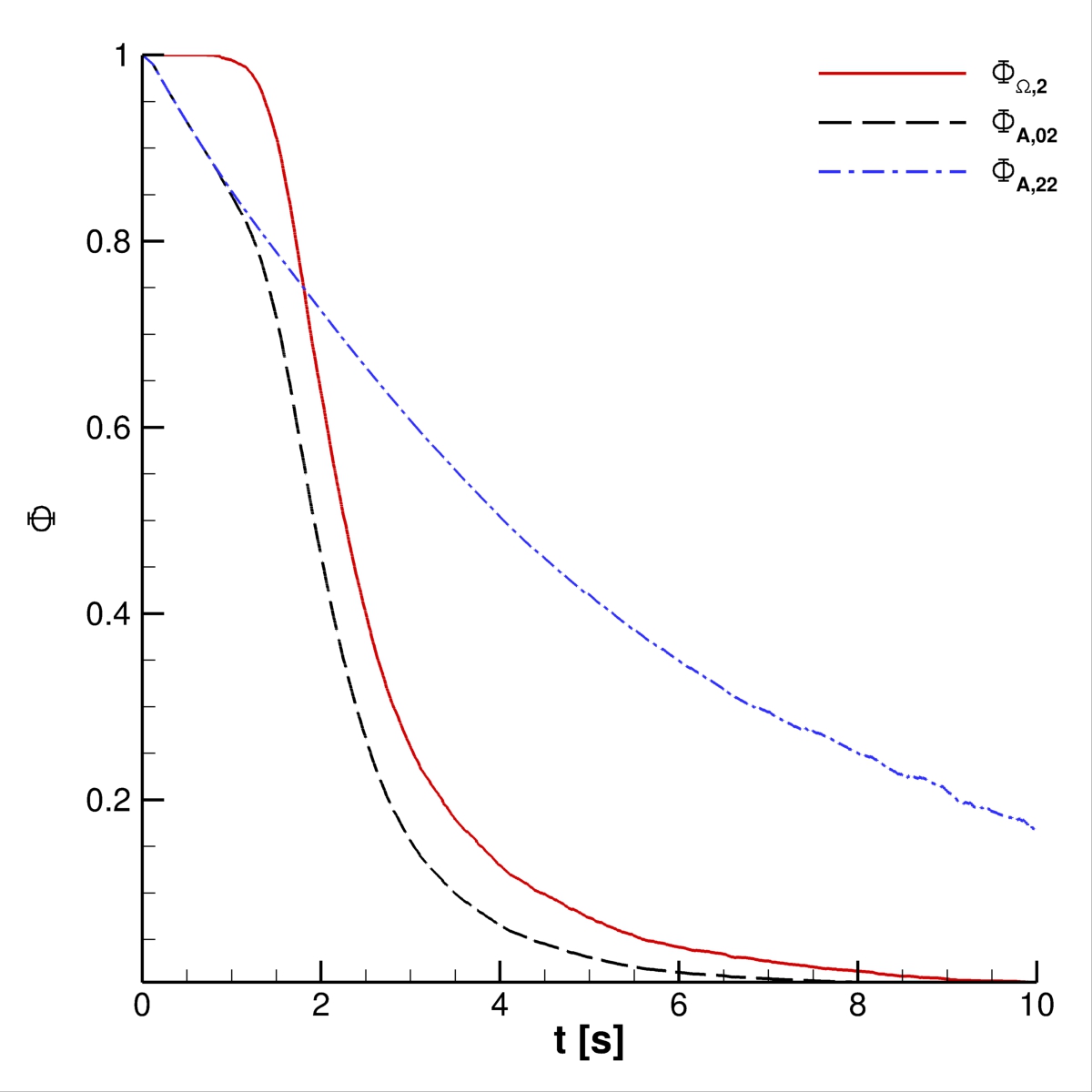}\label{fig:co_lpl}}
\caption{Effect of UV--C radiation for one cough ejection.}
\label{fig:uvc_cough1}
\end{figure}

\begin{figure}[htbp]
 \centering
 {\includegraphics[width=0.45\textwidth]{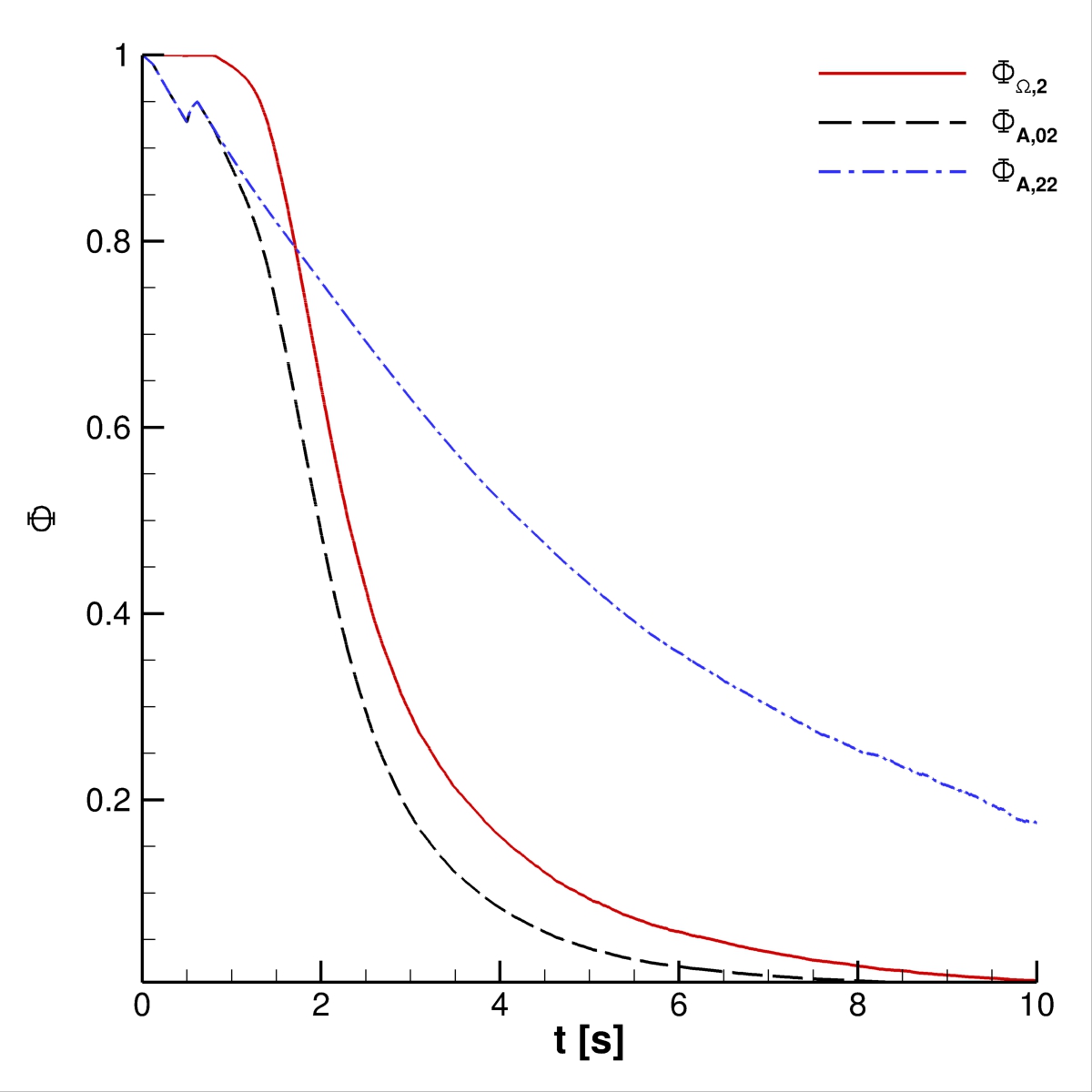}\label{fig:co_lpl}}
\caption{Effect of UV--C radiation for two cough ejections.}
\label{fig:uvc_cough2}
\end{figure}

\begin{figure}[htbp]
 \centering
 {\includegraphics[width=0.45\textwidth]{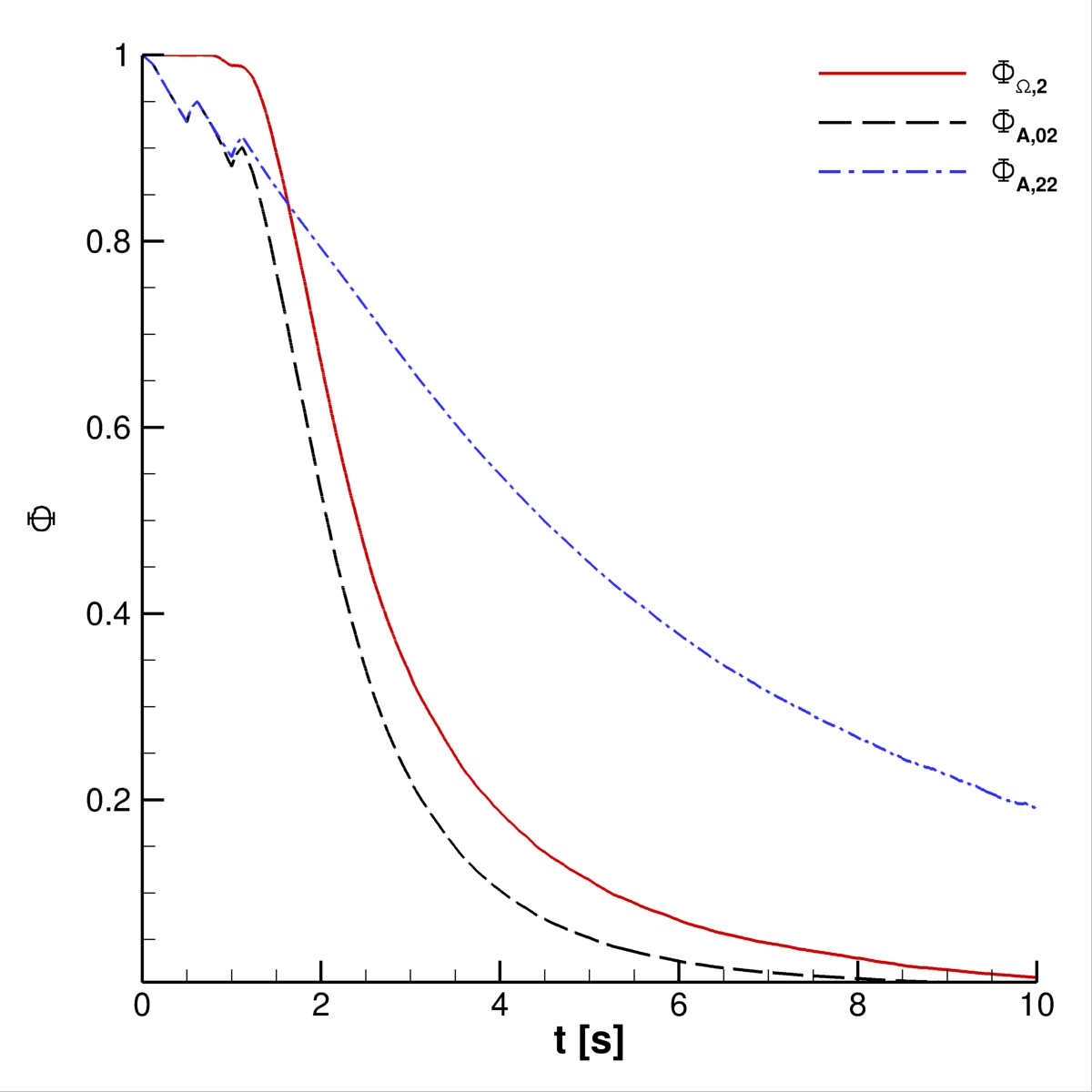}\label{fig:co_lpl}}
\caption{Effect of UV--C radiation for three cough ejections.}
\label{fig:uvc_cough3}
\end{figure}

The effect of UV--C lamp and orientation is clearly underlined in Fig.~\ref{fig:uvc_pow}. The $\Phi_{A,00}$ behaviour vis--\`a--vis time
is evident for the reported cases. Anyway, the impact of the lamp orientation is barely noticeable; the lower power level lamp
produces, according to the authors' opinion, a modest biological inactivation of parcels. Thus, hereinafter the reference configuration
uses a crossflow $55$ $\mathrm{W}$ lamp. 
The UV--C radiation inactivation capabilities are shown in Figs.~\ref{fig:uvc_cough1}--\ref{fig:uvc_cough3}.
Only $\Omega_{2}$ volume is studied since it is almost equivalent to $\Omega_{3}$  and $\Omega_{4}$ as regards
the droplets' presence.
It is very important to stress that permanence time of droplets has $10$ $\mathrm{s}$ scale for the different
coughing. This condition can be also noted observing Fig.~\ref{fig:cloud_d2}. Furthermore, UV--C light has a very good impact on the reduction of active particles. 
Indeed, $\Phi_{A,22}$ index rapidly decreases in time reducing the contamination risk.
In this context, it is interesting to note that after $4$ $\mathrm{s}$ the
number of parcels in $\Omega_{2}$ is low and they are mainly located in its bottom side where UV--C field intensity suddenly
decays, see Fig.~\ref{fig:cloud_d}.
Therefore, the overall effect of this technique can be considered very promising.

\section{Conclusions}\label{sec:concl}

This paper addresses the development and application of an Eulerian--Lagrangian
model for saliva droplets' cloud deriving from coughing. 
Particular emphasis has been devoted to SARS--CoV--2 biological inactivation
produced by UV--C radiation at $254$ $\mathrm{nm}$. Therefore, a new approach to 
to model the UV--C inactivation effect was introduced.\\ 
A preliminary work was dedicated to the correct evaluation of grid points spacing
as well as $\mathrm{Co_{max}}$ and parcels' injection features. Successively, 
several relevant configurations have been analysed in absence of an external wind.
Moreover, two indexes were introduced
in order to provide quantitative informations about the contamination risk.\\
Firstable we noted that only a few particles get to distances grater than $1.0$ $\mathrm{m}$,
in the vertical range $1.3 \mathrm{m} \le z \le 1.8 \mathrm{m}$, after different breathing conditions.
Hence, it may be confirmed that social distances greater than $1.0$ $\mathrm{m}$ do not reduce
notably the possibility of receiving infected particles in not windy conditions.
Furthermore, it was also showed that UV--C radiation is a promising technique to perform a real--time 
disinfection of a cloud resulting by coughing. 
Indeed, the number of active particles can be significantly reduced during the saliva droplets'
cloud evolution without giving a critical UV--C dose to the emitter.

\begin{acknowledgments}
The authors want to acknowledge Associazione Nazionale Big Data that awarded this research
work within COVID19--Fast access to HPC supercomputing facilities programme.
We acknowledge ENEA for awarding us access to CRESCO6  based at Portici. 
\end{acknowledgments}

\bibliography{aipsamp}

\begin{thebibliography}{33}%
\makeatletter
\providecommand \@ifxundefined [1]{%
 \@ifx{#1\undefined}
}%
\providecommand \@ifnum [1]{%
 \ifnum #1\expandafter \@firstoftwo
 \else \expandafter \@secondoftwo
 \fi
}%
\providecommand \@ifx [1]{%
 \ifx #1\expandafter \@firstoftwo
 \else \expandafter \@secondoftwo
 \fi
}%
\providecommand \natexlab [1]{#1}%
\providecommand \enquote  [1]{``#1''}%
\providecommand \bibnamefont  [1]{#1}%
\providecommand \bibfnamefont [1]{#1}%
\providecommand \citenamefont [1]{#1}%
\providecommand \href@noop [0]{\@secondoftwo}%
\providecommand \href [0]{\begingroup \@sanitize@url \@href}%
\providecommand \@href[1]{\@@startlink{#1}\@@href}%
\providecommand \@@href[1]{\endgroup#1\@@endlink}%
\providecommand \@sanitize@url [0]{\catcode `\\12\catcode `\$12\catcode
  `\&12\catcode `\#12\catcode `\^12\catcode `\_12\catcode `\%12\relax}%
\providecommand \@@startlink[1]{}%
\providecommand \@@endlink[0]{}%
\providecommand \url  [0]{\begingroup\@sanitize@url \@url }%
\providecommand \@url [1]{\endgroup\@href {#1}{\urlprefix }}%
\providecommand \urlprefix  [0]{URL }%
\providecommand \Eprint [0]{\href }%
\providecommand \doibase [0]{http://dx.doi.org/}%
\providecommand \selectlanguage [0]{\@gobble}%
\providecommand \bibinfo  [0]{\@secondoftwo}%
\providecommand \bibfield  [0]{\@secondoftwo}%
\providecommand \translation [1]{[#1]}%
\providecommand \BibitemOpen [0]{}%
\providecommand \bibitemStop [0]{}%
\providecommand \bibitemNoStop [0]{.\EOS\space}%
\providecommand \EOS [0]{\spacefactor3000\relax}%
\providecommand \BibitemShut  [1]{\csname bibitem#1\endcsname}%
\let\auto@bib@innerbib\@empty
\bibitem [{\citenamefont {Siegel}\ \emph {et~al.}(2007)\citenamefont {Siegel},
  \citenamefont {Rhinehart}, \citenamefont {Jackson},\ and\ \citenamefont
  {Chiarello}}]{Siegel:2007}%
  \BibitemOpen
  \bibfield  {author} {\bibinfo {author} {\bibfnamefont {J.}~\bibnamefont
  {Siegel}}, \bibinfo {author} {\bibfnamefont {E.}~\bibnamefont {Rhinehart}},
  \bibinfo {author} {\bibfnamefont {M.}~\bibnamefont {Jackson}}, \ and\
  \bibinfo {author} {\bibfnamefont {L.}~\bibnamefont {Chiarello}},\ }\href@noop
  {} {\enquote {\bibinfo {title} {{Guideline for Isolation Precautions:
  Preventing Transmission of Infectious Agents in Healthcare Settings}},}\ }
  (\bibinfo {year} {2007}),\ \bibinfo {note} {the Healthcare Infection Control
  Practices Advisory Committee, Atlanta, GA: U.S. Department of Health and
  Human Services}\BibitemShut {NoStop}%
\bibitem [{\citenamefont {Mittal}, \citenamefont {Ni},\ and\ \citenamefont
  {Seo}(2020)}]{Mittal2020}%
  \BibitemOpen
  \bibfield  {author} {\bibinfo {author} {\bibfnamefont {R.}~\bibnamefont
  {Mittal}}, \bibinfo {author} {\bibfnamefont {R.}~\bibnamefont {Ni}}, \ and\
  \bibinfo {author} {\bibfnamefont {J.-H.}\ \bibnamefont {Seo}},\ }\bibfield
  {title} {\enquote {\bibinfo {title} {{The flow physics of COVID-19}},}\
  }\href {\doibase 10.1017/jfm.2020.330} {\bibfield  {journal} {\bibinfo
  {journal} {Journal of Fluid Mechanics}\ }\textbf {\bibinfo {volume} {894}}
  (\bibinfo {year} {2020}),\ 10.1017/jfm.2020.330}\BibitemShut {NoStop}%
\bibitem [{\citenamefont {Ai}\ and\ \citenamefont {A.K.}(2018)}]{Melikov:2018}%
  \BibitemOpen
  \bibfield  {author} {\bibinfo {author} {\bibfnamefont {Z.}~\bibnamefont
  {Ai}}\ and\ \bibinfo {author} {\bibfnamefont {M.}~\bibnamefont {A.K.}},\
  }\bibfield  {title} {\enquote {\bibinfo {title} {Airborne spread of
  expiratory droplet nuclei between the occupants of indoor environments: A
  review},}\ }\href@noop {} {\bibfield  {journal} {\bibinfo  {journal}
  {Indoor}\ }\textbf {\bibinfo {volume} {28}},\ \bibinfo {pages} {500--524}
  (\bibinfo {year} {2018})}\BibitemShut {NoStop}%
\bibitem [{\citenamefont {Licina}\ \emph {et~al.}(2014)\citenamefont {Licina},
  \citenamefont {Pantelic}, \citenamefont {Melikov}, \citenamefont {Sekhar},\
  and\ \citenamefont {Tham}}]{LICINA}%
  \BibitemOpen
  \bibfield  {author} {\bibinfo {author} {\bibfnamefont {D.}~\bibnamefont
  {Licina}}, \bibinfo {author} {\bibfnamefont {J.}~\bibnamefont {Pantelic}},
  \bibinfo {author} {\bibfnamefont {A.}~\bibnamefont {Melikov}}, \bibinfo
  {author} {\bibfnamefont {C.}~\bibnamefont {Sekhar}}, \ and\ \bibinfo {author}
  {\bibfnamefont {K.}~\bibnamefont {Tham}},\ }\bibfield  {title} {\enquote
  {\bibinfo {title} {{Experimental investigation of the human convective
  boundary layer in a quiescent indoor environment}},}\ }\href@noop {}
  {\bibfield  {journal} {\bibinfo  {journal} {Building and Environment}\
  }\textbf {\bibinfo {volume} {75}},\ \bibinfo {pages} {79 -- 91} (\bibinfo
  {year} {2014})}\BibitemShut {NoStop}%
\bibitem [{\citenamefont {Nielsen}(2015)}]{NIELSEN}%
  \BibitemOpen
  \bibfield  {author} {\bibinfo {author} {\bibfnamefont {P.}~\bibnamefont
  {Nielsen}},\ }\bibfield  {title} {\enquote {\bibinfo {title} {{Fifty years of
  CFD for room air distribution}},}\ }\href@noop {} {\bibfield  {journal}
  {\bibinfo  {journal} {Building and Environment}\ }\textbf {\bibinfo {volume}
  {91}},\ \bibinfo {pages} {78 -- 90} (\bibinfo {year} {2015})},\ \bibinfo
  {note} {fifty Year Anniversary for Building and Environment}\BibitemShut
  {NoStop}%
\bibitem [{\citenamefont {Vuorinen}\ \emph {et~al.}(2020)\citenamefont
  {Vuorinen}, \citenamefont {Aarnio}, \citenamefont {Alava}, \citenamefont
  {Alopaeus}, \citenamefont {Atanasova}, \citenamefont {Auvinen}, \citenamefont
  {Balasubramanian}, \citenamefont {H.}, \citenamefont {Erasto}, \citenamefont
  {Grande}, \citenamefont {Hayward}, \citenamefont {Hellsten}, \citenamefont
  {Hostikka}, \citenamefont {Hokkanen}, \citenamefont {Kaario}, \citenamefont
  {Karvinen}, \citenamefont {Kivisto}, \citenamefont {Korhonen}, \citenamefont
  {Kosonen}, \citenamefont {Kuusela}, \citenamefont {Lestinen}, \citenamefont
  {Laurila}, \citenamefont {Nieminen}, \citenamefont {Peltonen}, \citenamefont
  {Pokki}, \citenamefont {Puisto}, \citenamefont {Raback}, \citenamefont
  {Salmenjoki}, \citenamefont {Sironen},\ and\ \citenamefont
  {Osterberg}}]{VUORINEN2020104866}%
  \BibitemOpen
  \bibfield  {author} {\bibinfo {author} {\bibfnamefont {V.}~\bibnamefont
  {Vuorinen}}, \bibinfo {author} {\bibfnamefont {M.}~\bibnamefont {Aarnio}},
  \bibinfo {author} {\bibfnamefont {M.}~\bibnamefont {Alava}}, \bibinfo
  {author} {\bibfnamefont {M.}~\bibnamefont {Alopaeus}}, \bibinfo {author}
  {\bibfnamefont {N.}~\bibnamefont {Atanasova}}, \bibinfo {author}
  {\bibfnamefont {M.}~\bibnamefont {Auvinen}}, \bibinfo {author} {\bibfnamefont
  {N.}~\bibnamefont {Balasubramanian}}, \bibinfo {author} {\bibfnamefont
  {B.}~\bibnamefont {H.}}, \bibinfo {author} {\bibfnamefont {P.}~\bibnamefont
  {Erasto}}, \bibinfo {author} {\bibfnamefont {R.}~\bibnamefont {Grande}},
  \bibinfo {author} {\bibfnamefont {N.}~\bibnamefont {Hayward}}, \bibinfo
  {author} {\bibfnamefont {A.}~\bibnamefont {Hellsten}}, \bibinfo {author}
  {\bibfnamefont {S.}~\bibnamefont {Hostikka}}, \bibinfo {author}
  {\bibfnamefont {J.}~\bibnamefont {Hokkanen}}, \bibinfo {author}
  {\bibfnamefont {O.}~\bibnamefont {Kaario}}, \bibinfo {author} {\bibfnamefont
  {A.}~\bibnamefont {Karvinen}}, \bibinfo {author} {\bibfnamefont
  {I.}~\bibnamefont {Kivisto}}, \bibinfo {author} {\bibfnamefont
  {M.}~\bibnamefont {Korhonen}}, \bibinfo {author} {\bibfnamefont
  {R.}~\bibnamefont {Kosonen}}, \bibinfo {author} {\bibfnamefont
  {J.}~\bibnamefont {Kuusela}}, \bibinfo {author} {\bibfnamefont
  {S.}~\bibnamefont {Lestinen}}, \bibinfo {author} {\bibfnamefont
  {E.}~\bibnamefont {Laurila}}, \bibinfo {author} {\bibfnamefont
  {H.}~\bibnamefont {Nieminen}}, \bibinfo {author} {\bibfnamefont
  {P.}~\bibnamefont {Peltonen}}, \bibinfo {author} {\bibfnamefont
  {J.}~\bibnamefont {Pokki}}, \bibinfo {author} {\bibfnamefont
  {A.}~\bibnamefont {Puisto}}, \bibinfo {author} {\bibfnamefont
  {P.}~\bibnamefont {Raback}}, \bibinfo {author} {\bibfnamefont
  {H.}~\bibnamefont {Salmenjoki}}, \bibinfo {author} {\bibfnamefont
  {T.}~\bibnamefont {Sironen}}, \ and\ \bibinfo {author} {\bibfnamefont
  {M.}~\bibnamefont {Osterberg}},\ }\bibfield  {title} {\enquote {\bibinfo
  {title} {{Modelling aerosol transport and virus exposure with numerical
  simulations in relation to SARS-CoV-2 transmission by inhalation indoors}},}\
  }\href@noop {} {\bibfield  {journal} {\bibinfo  {journal} {Safety Science}\
  }\textbf {\bibinfo {volume} {130}},\ \bibinfo {pages} {104866} (\bibinfo
  {year} {2020})}\BibitemShut {NoStop}%
\bibitem [{\citenamefont {Pendar}\ and\ \citenamefont {Pascoa}(2020)}]{Pendar}%
  \BibitemOpen
  \bibfield  {author} {\bibinfo {author} {\bibfnamefont {M.}~\bibnamefont
  {Pendar}}\ and\ \bibinfo {author} {\bibfnamefont {J.}~\bibnamefont
  {Pascoa}},\ }\bibfield  {title} {\enquote {\bibinfo {title} {Numerical
  modeling of the distribution of virus carrying saliva droplets during sneeze
  and cough},}\ }\href@noop {} {\bibfield  {journal} {\bibinfo  {journal}
  {Physics of Fluids}\ }\textbf {\bibinfo {volume} {32}},\ \bibinfo {pages}
  {083305} (\bibinfo {year} {2020})}\BibitemShut {NoStop}%
\bibitem [{\citenamefont {Dbouk}\ and\ \citenamefont
  {Drikakis}(2020{\natexlab{a}})}]{Dbouk}%
  \BibitemOpen
  \bibfield  {author} {\bibinfo {author} {\bibfnamefont {T.}~\bibnamefont
  {Dbouk}}\ and\ \bibinfo {author} {\bibfnamefont {D.}~\bibnamefont
  {Drikakis}},\ }\bibfield  {title} {\enquote {\bibinfo {title} {On coughing
  and airborne droplet transmission to humans},}\ }\href {\doibase
  10.1063/5.0011960} {\bibfield  {journal} {\bibinfo  {journal} {Physics of
  Fluids}\ }\textbf {\bibinfo {volume} {32}} (\bibinfo {year}
  {2020}{\natexlab{a}}),\ 10.1063/5.0011960}\BibitemShut {NoStop}%
\bibitem [{\citenamefont {Dbouk}\ and\ \citenamefont
  {Drikakis}(2020{\natexlab{b}})}]{Dbouk1}%
  \BibitemOpen
  \bibfield  {author} {\bibinfo {author} {\bibfnamefont {T.}~\bibnamefont
  {Dbouk}}\ and\ \bibinfo {author} {\bibfnamefont {D.}~\bibnamefont
  {Drikakis}},\ }\bibfield  {title} {\enquote {\bibinfo {title} {On respiratory
  droplets and face masks},}\ }\href@noop {} {\bibfield  {journal} {\bibinfo
  {journal} {Physics of Fluids}\ }\textbf {\bibinfo {volume} {32}},\ \bibinfo
  {pages} {063303} (\bibinfo {year} {2020}{\natexlab{b}})}\BibitemShut
  {NoStop}%
\bibitem [{\citenamefont {Dbouk}\ and\ \citenamefont
  {Drikakis}(2020{\natexlab{c}})}]{Dbouk2}%
  \BibitemOpen
  \bibfield  {author} {\bibinfo {author} {\bibfnamefont {T.}~\bibnamefont
  {Dbouk}}\ and\ \bibinfo {author} {\bibfnamefont {D.}~\bibnamefont
  {Drikakis}},\ }\bibfield  {title} {\enquote {\bibinfo {title} {Weather impact
  on airborne coronavirus survival},}\ }\href@noop {} {\bibfield  {journal}
  {\bibinfo  {journal} {Physics of Fluids}\ }\textbf {\bibinfo {volume} {32}},\
  \bibinfo {pages} {093312} (\bibinfo {year} {2020}{\natexlab{c}})}\BibitemShut
  {NoStop}%
\bibitem [{\citenamefont {Busco}\ \emph {et~al.}(2020)\citenamefont {Busco},
  \citenamefont {Yang}, \citenamefont {Seo},\ and\ \citenamefont
  {Hassan}}]{Busco}%
  \BibitemOpen
  \bibfield  {author} {\bibinfo {author} {\bibfnamefont {G.}~\bibnamefont
  {Busco}}, \bibinfo {author} {\bibfnamefont {S.}~\bibnamefont {Yang}},
  \bibinfo {author} {\bibfnamefont {J.}~\bibnamefont {Seo}}, \ and\ \bibinfo
  {author} {\bibfnamefont {Y.~A.}\ \bibnamefont {Hassan}},\ }\bibfield  {title}
  {\enquote {\bibinfo {title} {Sneezing and asymptomatic virus transmission},}\
  }\href@noop {} {\bibfield  {journal} {\bibinfo  {journal} {Physics of
  Fluids}\ }\textbf {\bibinfo {volume} {32}},\ \bibinfo {pages} {073309}
  (\bibinfo {year} {2020})}\BibitemShut {NoStop}%
\bibitem [{\citenamefont {{Abuhegazy, M. and Talaat, K. and Anderoglu, O. and
  Poroseva, S. V. }}(2020)}]{Abu_pof}%
  \BibitemOpen
  \bibfield  {author} {\bibinfo {author} {\bibnamefont {{Abuhegazy, M. and
  Talaat, K. and Anderoglu, O. and Poroseva, S. V. }}},\ }\bibfield  {title}
  {\enquote {\bibinfo {title} {{Numerical investigation of aerosol transport in
  a classroom with relevance to COVID--19}},}\ }\href@noop {} {\bibfield
  {journal} {\bibinfo  {journal} {Physics of Fluids}\ }\textbf {\bibinfo
  {volume} {32}},\ \bibinfo {pages} {103311} (\bibinfo {year}
  {2020})}\BibitemShut {NoStop}%
\bibitem [{\citenamefont {{Li,H. and Leong, F.Y. and Xu, G. and Ge, Z. and
  Kang,C.W. and Lim, K.H. }}(2020)}]{Li_pof}%
  \BibitemOpen
  \bibfield  {author} {\bibinfo {author} {\bibnamefont {{Li,H. and Leong, F.Y.
  and Xu, G. and Ge, Z. and Kang,C.W. and Lim, K.H. }}},\ }\bibfield  {title}
  {\enquote {\bibinfo {title} {{Dispersion of evaporating cough droplets in
  tropical outdoor environment}},}\ }\href {\doibase 10.1063/5.0026360}
  {\bibfield  {journal} {\bibinfo  {journal} {Physics of Fluids}\ }\textbf
  {\bibinfo {volume} {32}},\ \bibinfo {pages} {113301} (\bibinfo {year}
  {2020})}\BibitemShut {NoStop}%
\bibitem [{\citenamefont {Weller}\ \emph {et~al.}(1998)\citenamefont {Weller},
  \citenamefont {Tabor}, \citenamefont {Jasak},\ and\ \citenamefont
  {Fureby}}]{OF_paper}%
  \BibitemOpen
  \bibfield  {author} {\bibinfo {author} {\bibfnamefont {H.}~\bibnamefont
  {Weller}}, \bibinfo {author} {\bibfnamefont {G.}~\bibnamefont {Tabor}},
  \bibinfo {author} {\bibfnamefont {H.}~\bibnamefont {Jasak}}, \ and\ \bibinfo
  {author} {\bibfnamefont {C.}~\bibnamefont {Fureby}},\ }\bibfield  {title}
  {\enquote {\bibinfo {title} {{A tensorial approach to computational continuum
  mechanics using object--oriented techniques}},}\ }\href@noop {} {\bibfield
  {journal} {\bibinfo  {journal} {Computational Physics}\ }\textbf {\bibinfo
  {volume} {12}},\ \bibinfo {pages} {620--631} (\bibinfo {year}
  {1998})}\BibitemShut {NoStop}%
\bibitem [{\citenamefont {Buchan}, \citenamefont {Yang},\ and\ \citenamefont
  {Atkinson}(2020)}]{Buchan2020}%
  \BibitemOpen
  \bibfield  {author} {\bibinfo {author} {\bibfnamefont {A.}~\bibnamefont
  {Buchan}}, \bibinfo {author} {\bibfnamefont {L.}~\bibnamefont {Yang}}, \ and\
  \bibinfo {author} {\bibfnamefont {K.~D.}\ \bibnamefont {Atkinson}},\
  }\bibfield  {title} {\enquote {\bibinfo {title} {{Predicting airborne
  coronavirus inactivation by far-UVC in populated rooms using a high-fidelity
  coupled radiation-CFD model}},}\ }\href@noop {} {\bibfield  {journal}
  {\bibinfo  {journal} {Scientific Reports}\ }\textbf {\bibinfo {volume} {10}}
  (\bibinfo {year} {2020})}\BibitemShut {NoStop}%
\bibitem [{\citenamefont {Menter}(1994)}]{Menter.1994}%
  \BibitemOpen
  \bibfield  {author} {\bibinfo {author} {\bibfnamefont {F.}~\bibnamefont
  {Menter}},\ }\bibfield  {title} {\enquote {\bibinfo {title} {{Two-Equation
  Eddy Viscosity Turbulence Models for Engineering Applications}},}\
  }\href@noop {} {\bibfield  {journal} {\bibinfo  {journal} {AIAA Journal}\
  }\textbf {\bibinfo {volume} {32}},\ \bibinfo {pages} {1598 -- 1695} (\bibinfo
  {year} {1994})}\BibitemShut {NoStop}%
\bibitem [{\citenamefont {Crowe}, \citenamefont {Sharma},\ and\ \citenamefont
  {Stock}(1977)}]{Crowe:1977}%
  \BibitemOpen
  \bibfield  {author} {\bibinfo {author} {\bibfnamefont {C.~T.}\ \bibnamefont
  {Crowe}}, \bibinfo {author} {\bibfnamefont {M.~P.}\ \bibnamefont {Sharma}}, \
  and\ \bibinfo {author} {\bibfnamefont {D.~E.}\ \bibnamefont {Stock}},\
  }\bibfield  {title} {\enquote {\bibinfo {title} {{The Particle-Source-In Cell
  (PSI-CELL) Model for Gas-Droplet Flows}},}\ }\href@noop {} {\bibfield
  {journal} {\bibinfo  {journal} {Journal of Fluids Engineering}\ }\textbf
  {\bibinfo {volume} {99}},\ \bibinfo {pages} {325--332} (\bibinfo {year}
  {1977})}\BibitemShut {NoStop}%
\bibitem [{\citenamefont {Putnam}(1961)}]{Putnam:1961}%
  \BibitemOpen
  \bibfield  {author} {\bibinfo {author} {\bibfnamefont {A.}~\bibnamefont
  {Putnam}},\ }\bibfield  {title} {\enquote {\bibinfo {title} {Integratable
  form of droplet drag coefficient},}\ }\href@noop {} {\bibfield  {journal}
  {\bibinfo  {journal} {ARS J.}\ }\textbf {\bibinfo {volume} {31}},\ \bibinfo
  {pages} {1467 -- 1468} (\bibinfo {year} {1961})}\BibitemShut {NoStop}%
\bibitem [{\citenamefont {{Longest, P.W. and Jinxiang Xi, J.
  }}(2007)}]{aerosol:2007}%
  \BibitemOpen
  \bibfield  {author} {\bibinfo {author} {\bibnamefont {{Longest, P.W. and
  Jinxiang Xi, J. }}},\ }\bibfield  {title} {\enquote {\bibinfo {title}
  {{Effectiveness of Direct Lagrangian Tracking Models for Simulating
  Nanoparticle Deposition in the Upper Airways}},}\ }\href@noop {} {\bibfield
  {journal} {\bibinfo  {journal} {Aerosol Science and Technology}\ }\textbf
  {\bibinfo {volume} {41}},\ \bibinfo {pages} {380--397} (\bibinfo {year}
  {2007})}\BibitemShut {NoStop}%
\bibitem [{\citenamefont {{Zhao, B. and Zhang, Y. and Li, X. and Yang, X. and
  Huang, D.}}(2004)}]{ZHAO20041}%
  \BibitemOpen
  \bibfield  {author} {\bibinfo {author} {\bibnamefont {{Zhao, B. and Zhang, Y.
  and Li, X. and Yang, X. and Huang, D.}}},\ }\bibfield  {title} {\enquote
  {\bibinfo {title} {{Comparison of indoor aerosol particle concentration and
  deposition in different ventilated rooms by numerical method}},}\ }\href@noop
  {} {\bibfield  {journal} {\bibinfo  {journal} {Building and Environment}\
  }\textbf {\bibinfo {volume} {39}},\ \bibinfo {pages} {1 -- 8} (\bibinfo
  {year} {2004})}\BibitemShut {NoStop}%
\bibitem [{\citenamefont {Ranz}\ and\ \citenamefont
  {Marshall}(1952)}]{Ranz:1952}%
  \BibitemOpen
  \bibfield  {author} {\bibinfo {author} {\bibfnamefont {W.~E.}\ \bibnamefont
  {Ranz}}\ and\ \bibinfo {author} {\bibfnamefont {W.~R.}\ \bibnamefont
  {Marshall}},\ }\bibfield  {title} {\enquote {\bibinfo {title} {Evaporation
  from drops},}\ }\href@noop {} {\bibfield  {journal} {\bibinfo  {journal}
  {Chem. Eng. Prog.}\ }\textbf {\bibinfo {volume} {48}},\ \bibinfo {pages} {141
  -- 146} (\bibinfo {year} {1952})}\BibitemShut {NoStop}%
\bibitem [{\citenamefont {Mugele}\ and\ \citenamefont
  {Evans}(1951)}]{Mugele1951}%
  \BibitemOpen
  \bibfield  {author} {\bibinfo {author} {\bibfnamefont {R.~A.}\ \bibnamefont
  {Mugele}}\ and\ \bibinfo {author} {\bibfnamefont {H.~D.}\ \bibnamefont
  {Evans}},\ }\bibfield  {title} {\enquote {\bibinfo {title} {{Droplet Size
  Distribution in Sprays}},}\ }\href@noop {} {\bibfield  {journal} {\bibinfo
  {journal} {{Industrial {\&} Engineering Chemistry}}\ }\textbf {\bibinfo
  {volume} {43}},\ \bibinfo {pages} {1317--1324} (\bibinfo {year}
  {1951})}\BibitemShut {NoStop}%
\bibitem [{\citenamefont {Xie}\ \emph {et~al.}(2012)\citenamefont {Xie},
  \citenamefont {Li}, \citenamefont {Sun},\ and\ \citenamefont
  {Liu}}]{Xie:2009}%
  \BibitemOpen
  \bibfield  {author} {\bibinfo {author} {\bibfnamefont {X.}~\bibnamefont
  {Xie}}, \bibinfo {author} {\bibfnamefont {Y.}~\bibnamefont {Li}}, \bibinfo
  {author} {\bibfnamefont {H.}~\bibnamefont {Sun}}, \ and\ \bibinfo {author}
  {\bibfnamefont {L.}~\bibnamefont {Liu}},\ }\bibfield  {title} {\enquote
  {\bibinfo {title} {{Exhaled droplets due to talking and coughing}},}\
  }\href@noop {} {\bibfield  {journal} {\bibinfo  {journal} {J R Soc
  Interface}\ } (\bibinfo {year} {2012})}\BibitemShut {NoStop}%
\bibitem [{\citenamefont {{Beggs, C.B. and Kerr, K.G. and Donnelly, J.K. and
  Sleigh, P.A. and Mara, D.D. and Cairns, G.}}(2000)}]{Beggs2000141}%
  \BibitemOpen
  \bibfield  {author} {\bibinfo {author} {\bibnamefont {{Beggs, C.B. and Kerr,
  K.G. and Donnelly, J.K. and Sleigh, P.A. and Mara, D.D. and Cairns, G.}}},\
  }\bibfield  {title} {\enquote {\bibinfo {title} {{An engineering approach to
  the control of Mycobacterium tuberculosis and other airborne pathogens: A UK
  hospital based pilot study}},}\ }\href {\doibase
  10.1016/S0035-9203(00)90250-5} {\bibfield  {journal} {\bibinfo  {journal}
  {{Transactions of the Royal Society of Tropical Medicine and Hygiene}}\
  }\textbf {\bibinfo {volume} {94}},\ \bibinfo {pages} {141--146} (\bibinfo
  {year} {2000})}\BibitemShut {NoStop}%
\bibitem [{\citenamefont {{C.J. Noakes and L.A. Fletcher and C.B. Beggs and
  P.A. Sleigh and K.G. Kerr}}(2004)}]{NOAKES2004489}%
  \BibitemOpen
  \bibfield  {author} {\bibinfo {author} {\bibnamefont {{C.J. Noakes and L.A.
  Fletcher and C.B. Beggs and P.A. Sleigh and K.G. Kerr}}},\ }\bibfield
  {title} {\enquote {\bibinfo {title} {{Development of a numerical model to
  simulate the biological inactivation of airborne microorganisms in the
  presence of ultraviolet light}},}\ }\href@noop {} {\bibfield  {journal}
  {\bibinfo  {journal} {Journal of Aerosol Science}\ }\textbf {\bibinfo
  {volume} {35}},\ \bibinfo {pages} {489 -- 507} (\bibinfo {year}
  {2004})}\BibitemShut {NoStop}%
\bibitem [{\citenamefont {{Kowalski, W.J. Walsh, T.J. and Petraitis,
  V.}}(2020)}]{KowalskiUVC}%
  \BibitemOpen
  \bibfield  {author} {\bibinfo {author} {\bibnamefont {{Kowalski, W.J. Walsh,
  T.J. and Petraitis, V.}}},\ }\href@noop {} {\enquote {\bibinfo {title} {{2020
  COVID-19 Coronavirus ultraviolet susceptibility}},}\ } (\bibinfo {year}
  {2020})\BibitemShut {NoStop}%
\bibitem [{\citenamefont {Modest}(2014)}]{Modest}%
  \BibitemOpen
  \bibfield  {author} {\bibinfo {author} {\bibfnamefont {M.}~\bibnamefont
  {Modest}},\ }\href@noop {} {\emph {\bibinfo {title} {{Radiative Heat
  Transfer}}}}\ (\bibinfo  {publisher} {Elsevier},\ \bibinfo {address} {New
  York},\ \bibinfo {year} {2014})\BibitemShut {NoStop}%
\bibitem [{\citenamefont {Kowalski}\ \emph {et~al.}(2000)\citenamefont
  {Kowalski}, \citenamefont {Bahnfleth}, \citenamefont {Witham}, \citenamefont
  {Severin},\ and\ \citenamefont {Whittam}}]{Kowalski2000}%
  \BibitemOpen
  \bibfield  {author} {\bibinfo {author} {\bibfnamefont {W.~J.}\ \bibnamefont
  {Kowalski}}, \bibinfo {author} {\bibfnamefont {W.~P.}\ \bibnamefont
  {Bahnfleth}}, \bibinfo {author} {\bibfnamefont {D.~L.}\ \bibnamefont
  {Witham}}, \bibinfo {author} {\bibfnamefont {B.~F.}\ \bibnamefont {Severin}},
  \ and\ \bibinfo {author} {\bibfnamefont {T.~S.}\ \bibnamefont {Whittam}},\
  }\bibfield  {title} {\enquote {\bibinfo {title} {{Mathematical Modeling of
  Ultraviolet Germicidal Irradiation for Air Disinfection}},}\ }\href@noop {}
  {\bibfield  {journal} {\bibinfo  {journal} {{Quantitative Microbiology}}\
  }\textbf {\bibinfo {volume} {2}},\ \bibinfo {pages} {249--270} (\bibinfo
  {year} {2000})}\BibitemShut {NoStop}%
\bibitem [{\citenamefont {Issa}(1986)}]{Issa198640}%
  \BibitemOpen
  \bibfield  {author} {\bibinfo {author} {\bibfnamefont {R.}~\bibnamefont
  {Issa}},\ }\bibfield  {title} {\enquote {\bibinfo {title} {Solution of the
  implicitly discretised fluid flow equations by operator-splitting},}\
  }\href@noop {} {\bibfield  {journal} {\bibinfo  {journal} {Journal of
  Computational Physics}\ }\textbf {\bibinfo {volume} {62}},\ \bibinfo {pages}
  {40--65} (\bibinfo {year} {1986})}\BibitemShut {NoStop}%
\bibitem [{\citenamefont {{Scharfman, B. E. and Techet, A. H. and Bush, J. W.
  M. and Bourouiba, L.}}(2016)}]{Scharfman2016}%
  \BibitemOpen
  \bibfield  {author} {\bibinfo {author} {\bibnamefont {{Scharfman, B. E. and
  Techet, A. H. and Bush, J. W. M. and Bourouiba, L.}}},\ }\bibfield  {title}
  {\enquote {\bibinfo {title} {{Visualization of sneeze ejecta: steps of fluid
  fragmentation leading to respiratory droplets}},}\ }\href@noop {} {\bibfield
  {journal} {\bibinfo  {journal} {Experiments in Fluids}\ }\textbf {\bibinfo
  {volume} {57}},\ \bibinfo {pages} {24} (\bibinfo {year} {2016})}\BibitemShut
  {NoStop}%
\bibitem [{\citenamefont {{Van Der Reijden, W.A. and Veerman, E.C.I. and Nieuw
  Amerongen, A.V.}}(1993)}]{VanDerReijden1993141}%
  \BibitemOpen
  \bibfield  {author} {\bibinfo {author} {\bibnamefont {{Van Der Reijden, W.A.
  and Veerman, E.C.I. and Nieuw Amerongen, A.V.}}},\ }\bibfield  {title}
  {\enquote {\bibinfo {title} {{Shear rate dependent viscoelastic behavior of
  human glandular salivas}},}\ }\href {\doibase 10.3233/BIR-1993-30205}
  {\bibfield  {journal} {\bibinfo  {journal} {Biorheology}\ }\textbf {\bibinfo
  {volume} {30}},\ \bibinfo {pages} {141--152} (\bibinfo {year}
  {1993})}\BibitemShut {NoStop}%
\bibitem [{\citenamefont {{Sula, C. and Grosshans, H. and Papalexandris, M.
  V.}}(2020)}]{Sula2020}%
  \BibitemOpen
  \bibfield  {author} {\bibinfo {author} {\bibnamefont {{Sula, C. and
  Grosshans, H. and Papalexandris, M. V.}}},\ }\bibfield  {title} {\enquote
  {\bibinfo {title} {{Assessment of Droplet Breakup Models for Spray Flow
  Simulations}},}\ }\href@noop {} {\bibfield  {journal} {\bibinfo  {journal}
  {{Flow, Turbulence and Combustion}}\ }\textbf {\bibinfo {volume} {105}},\
  \bibinfo {pages} {889--914} (\bibinfo {year} {2020})}\BibitemShut {NoStop}%
\bibitem [{\citenamefont {{Navy Environmental Health
  Center--USA}}(1992)}]{UVCguide}%
  \BibitemOpen
  \bibfield  {author} {\bibinfo {author} {\bibnamefont {{Navy Environmental
  Health Center--USA}}},\ }\href@noop {} {\enquote {\bibinfo {title}
  {{Ultraviolet Radiation Guide}},}\ } (\bibinfo {year} {1992})\BibitemShut
  {NoStop}%
\end{thebibliography}%

\end{document}